\documentclass[fleqn,usenatbib]{mnras}

\usepackage{newtxtext,newtxmath}

\usepackage[T1]{fontenc}
\usepackage{amsmath}

\DeclareRobustCommand{\VAN}[3]{#2}
\let\VANthebibliography\thebibliography
\def\thebibliography{\DeclareRobustCommand{\VAN}[3]{##3}\VANthebibliography}

\usepackage[dvipsnames]{xcolor}
\usepackage{graphicx}	
\usepackage{amsmath}	

\newcommand{\msol}{M$_{\odot}$\,}

\title[Galaxy Formation as a Cosmological Tool]{The Revised Evolutionary Volume Tolman Test: Cosmological Constraints from Galaxy Evolution}

\author[]{{
Christopher J. Conselice,$^{1}$\thanks{E-mail: conselice@manchester.ac.uk} Edmund J. Copeland$^{2}$\thanks{E-mail: edmund.copeland@nottingham.ac.uk } and Sergio Sevillano Mu\~{n}oz},$^{2,3,4}$\thanks{E-mail: sergiosm@sas.upenn.edu }\\
$^{1}$Jodrell Bank Centre for Astrophysics, University of Manchester, Oxford Road, Manchester, UK \\
$^{2}$ School of Physics and Astronomy, University of Nottingham, University Park, Nottingham, NG7 2RD, UK\\
$^{3}$ Institute for Particle Physics Phenomenology, Department of Physics, Durham University, Durham, DH1 3LE, UK\\
$^{4}$ Center for Particle Cosmology, Department of Physics and Astronomy, University of Pennsylvania, Philadelphia, Pennsylvania 19104, USA
}

\date{Accepted XXX. Received YYY; in original form ZZZ}

\pubyear{2022}

\begin{document}
\label{firstpage}
\pagerange{\pageref{firstpage}--\pageref{lastpage}}
\maketitle

\begin{abstract}
In this study we adapt a classical cosmology measurement, the volume or number density test, to a modern synthesis of observed galaxy evolution. We do this by using measured galaxy mass functions and the history of galaxy evolution through star formation and galaxy mergers, inspired by the latest results from deep extragalactic surveys. We develop a new framework using measured galaxy volume number densities as a function of redshift and volume to determine cosmological parameters, especially those which alter the volume of the Universe at a given redshift. Whilst this is a classic cosmology test proposed since at least the 1930s, it has largely been abandoned for decades due to uncertainties in galaxy evolution which make it difficult to trace galaxy populations through time. However, recent advances in our understanding of star formation and the merging history of galaxies allow us to revise this method to uncover and measure cosmological parameters, especially those which involve the nature of dark energy.  We present a modified version of the volume test, called the revised evolutionary volume Tolman test, using properties of known galaxy evolution as part of the cosmological calculation. We show how this method can successfully be applied and is competitive with other major cosmological measurement methods, including those using SNe and the CMB, when the merger and star formation histories can be measured accurately to between 1–10\%. This accuracy is not yet achievable, but we discuss how future missions will allow these astrophysical quantities to be known at this level. Within this measurement accuracy we can measure the dynamical properties of dark energy, including its evolution through its equation of state. We also give a fuller accounting of the future use of this new method with upcoming galaxy surveys such as {\em Euclid} and LSST/Rubin.

\end{abstract}




\section{Introduction}

One of the primary goals in astrophysics is understanding the cosmological history of the universe which includes measuring the cosmological parameters determining this evolution.  Currently, one of the major questions within cosmology is the nature of the expansion of the universe, or the origin
of what is now called dark energy (For reviews of dark energy see \cite{Copeland:2006wr,Frieman:2008sn}).  Understanding and measuring the dark energy and other cosmological parameters is now a major industry in astronomy, which are the focus of cosmic microwave background (CMB) missions such as WMAP \citep{Larson:2010gs}, Planck \citep{Planck:2018vyg}, ACT \citep{ACT:2020gnv} as well as galaxy surveys such as DESI  \citep[][]{DESI:2024mwx}, DES and Euclid \citep[][]{Mellier2024} among others.
Meanwhile surveys of distant Cephid variables (\cite{Madore:1991yf}) and Type 1a supernova \citep{SupernovaSearchTeam:1998fmf,SupernovaCosmologyProject:1998vns} have led to independent determinations of the Hubble parameter (For a recent update see \citep{Freedman:2023jcz} and \citep{2025arXiv251023823H}), some of which differ from the CMB prediction and which has led to what has become known as the Hubble Tension (for a discussion of the tension and possible theoretical resolutions see \cite{Kamionkowski:2022pkx} and \cite{Knox:2019rjx}).   

Measuring the evolution of dark energy is a major goal of these ongoing and future missions, and a number of interesting and intriguing results are beginning to emerge. For example, using a combination of Baryon Acoustic Oscillation and Type 1a Supernova data, the DESI collaboration have argued for the first tentative evidence of dynamical dark energy \citep[e.g.,][]{DESI:2024mwx, DESI:2025zgx}, although this has met with some resistance, see for example \citep[e.g.,][]{Cortes:2024lgw, Cortes:2025joz, Efstathiou:2024xcq, Wolf:2025jed}. In the next
decade very large projects such as Euclid \citep{Amendola:2016saw, Mellier2024}, LSST/Rubin \citep{LSST:2008ijt} and WFIRST \citep{Hounsell:2017ejq}
will continue this quest to understand the nature of dark energy.  Most of these studies are done using traditional methods that have been heavily developed and built up over decades. This includes not only CMB power-spectra, but also examining the evolution of structure in the universe through weak lensing and galaxy clustering, as well as measuring the abundances of galaxy clusters over time \citep[e.g.,][]{DES2018}. 



Whereas the analysis of Cepheid variables and Type 1a SN in distant galaxies has led to the determination of dark energy through luminosity distance measurements, this is not the first time galaxies have been used in cosmology to determine the evolution of the universe. In fact historically, cosmology was founded on the use of galaxies as cosmological probes, although the methods have in the past few decades flittered out due to the large scatter in measurables \citep[e.g.,][]{Sandage1988}. Given that, galaxies as a cosmological probe is potentially an unexplored and long discarded method for uncovering cosmological features, which could help resolve some of the existing discrepancies by providing a new approach to determine cosmological parameters \citep[e.g.,][]{Loh:1986wg, 2000ApJ...534L..11N, Conselice2014morph}.


 Since its first proposal, galaxies have been recognised as a potentially powerful method for measuring cosmological properties.  In the 1930s a few pioneering papers, \citep[e.g.,][]{Tolman:1934za, HubbleTolman1935} introduced a novel test which made use of galaxies, namely through a number count method with magnitude.  This test was such that by determining the galaxy counts in a survey region down to an apparent magnitude, m, the total number of galaxies should grow as log $N \sim 0.6 \cdot {\rm log~m}$ in a Euclidean universe. By measuring the number of galaxies down to some magnitude the idea was to fit the slope between the magnitude and number counts.  One success of this method was to show that the steady-state model of the universe was not consistent with radio galaxy counts \citep[][]{ryle1961}.  In general, however, deviations from a $\sim 0.6$ slope would reveal if the universe had some positive or negative curvature. Since this method was proposed it was gradually realised that the number counts would depend on curvature of space, cosmological expansion, and the evolution of galaxies.  Only later was it considered a cosmological test, in principle, and began appearing in cosmology textbooks \citep[e.g.,][]{Peebles1980}, but it was rarely, or never fully, developed into a method of measurement. 

Later this method was expanded to (in principle) include the number of galaxies in a given cosmological volume in a given redshift bin, usually in terms of galaxy counts \citep[e.g.,][]{Shanks1984,Yoshii1988A, Cowie1991}.  This approach however was impossible to use to determine precise cosmological features as there was a degeneracy between the largely unknown galaxy distances, the fact of a 'blue' galaxy excess, and the cosmology itself. Often there was an excess of blue galaxies seen that could be due to galaxy evolution or due to certain values of the de-acceleration parameter, q$_0$ \citep[e.g.,][]{Shanks1984, Broadhurst}.  This approach was however largely abandoned once redshifts for these objects became available with the establishment of multi-wavelength deep imaging from the Hubble Space Telescope and the discovery and study of Lyman-break galaxies \citep[e.g.,][]{Steidel1996, Williams1996, Ellis1997}.

The method of looking at volume densities has thus never been fully developed for use with astrophysical processes in galaxies. In this way cosmological parameters can be inferred from the evolution of this number density across redshift bins with time.  The fact that galaxies evolve meant that by themselves they were eventually considered as poor standard candles hence could not be directly used to obtain a measure of distances to galaxies \citep[e.g.,][]{Tinsley1968}.

The use of a volume test for cosmology was reexamined, in relatively modern times, by \cite{Loh:1986wg} who used a version of the method to measure the total mass density parameter as $\Omega = 0.9_{-0.5}^{+0.6}$ using 1000 galaxies with a median
redshift of $z = 0.5$.  They assumed in their analysis that both the number density of galaxies and the form of the luminosity
function does not change with redshift.  Today we know that these assumptions are
incorrect, as galaxy mergers can have a profound effect on number densities \citep[e.g.,][]{1988ApJ...326L...1B, Conselice:2016zid}.  Furthermore, we
know that galaxy evolution occurs in a 'downsizing' pattern whereby the most massive galaxies end their star formation earlier than lower mass systems \citep[e.g.,][]{Bundy:2005kd, pacifici2016timing}.    In another relatively recent study \cite{2000ApJ...534L..11N} investigated ways to measure the cosmological equation of state with counts of galaxies using galaxy kinematics as a measure of a galaxy's mass.    Other attempts to use galaxy evolution for measuring csomology includes using the merger history for finding the equation of state of the universe \cite[e.g.,][]{Conselice2014morph}.

In this paper, we use the latest results from studies of galaxy formation and evolution to revisit and revise the classic test of \cite{Tolman:1934za}. 
Our goal is to determine the set of cosmological parameters that can be measured in the universe by tracing galaxy co-moving volumes, including how well these features can be measured.   The new twist we add to this classic measurement is to include the effects of galaxy evolution and formation which can change the number of galaxies as a function of redshift in an analogous manner to how cosmology might change the number \citep[e.g.,][]{Conselice:2016zid}. Then by removing the effects of galaxy evolution, we can in principle determine a 'clean' evolution of the number of galaxies with redshift.    

The basic idea we explore is using galaxies as a volume probe as a function of redshift for different cosmologies, and in particular to focus on what is considered today as realistic variations around the basic $\Lambda$CDM model.   Whilst this idea has existed for nearly a century, it has been largely abandoned due to the difficulty in understanding astrophysical effects which can mimic changes in galaxy counts in a volume as a function of redshift.  This is a non-trivial problem for many reasons however, and ultimately to be competitive with other well developed methods such as SNe and CMB measurements will require a very good knowledge of not only galaxy evolution
and formation, but also a reliable standard for measuring a standard galaxy scale that can be traced through time. We show that by calculating the galaxy number counts in volumes for a range of cosmologies, with varying types and forms of dark energy, 
it will be possible in the coming decade or so, to use this diagnostic as a powerful discriminator between cosmological models.

The outline of this paper is as follows: 
we begin in Section \ref{sec:theory} with an explanation of the theory behind the galaxy counts method. Section \ref{galaxy-star} gives an overview into the physics and the impact behind galaxy mergers and star formation, which directly affects the galaxy number counts. We show how the number density of galaxies is modified by the inclusion of star formation rates (which increases the number of galaxies), and galaxy merger rates (which can both increase and decrease them), emphasising the redshift dependence of these contributions. In section \ref{Sec:Results}, having determined the modified Schechter function we use it to compare four models of dark energy -- $\Lambda$CDM, and three dynamical dark energy models, showing that as the uncertainties in the galaxy dynamics reduces over time, we will eventually be able to differentiate between $\Lambda$CDM and different dynamical dark energy paradigms. Finally we discuss our results and conclude with possible improvements and future directions in Section \ref{sec:disc}. Throughout this paper, unless otherwise stated, we  assume a $\Lambda$CDM cosmology.  

\section{Method}
\label{sec:theory}

The method that we use here is essentially a modified version of  the classic volume test for cosmology first proposed by \cite{Tolman:1934za} and \cite{HubbleTolman1935} but later expanded and investigated by others \citep[e.g.,][]{Loh:1986wg, 2000ApJ...534L..11N}, although these studies never instigated any major follow up. Our approach here includes important differences from these earlier studies, making it distinct from these previous methods. While this was once considered as a possible direct cosmological probe, this method has largely been discarded for serious use due to the fact that galaxies evolve and therefore observed changes in galaxy counts could be due to either cosmological or galaxy evolution effects, but impossible to separate.  

Observations and our knowledge of galaxy astrophysics has now made it to the point where we can break this degeneracy and extract the cosmology in the volume test. This is due to the fact that studies of external galaxies has reached the stage where we can now measure a slew of properties that allow us to directly trace how galaxy formation occurs.  Galaxy evolution and formation has traditionally been measured through bulk properties such as stellar mass, but now we can measure them directly through active formation processes as seen through star formation and merging activity \cite[e.g.,][]{Conselice2014morph, Mundy:2015qgi, Ownsworth2016, Whitney2021, Duan2024}.  

\begin{figure*}
    \centering
    \includegraphics[width=\linewidth]
    {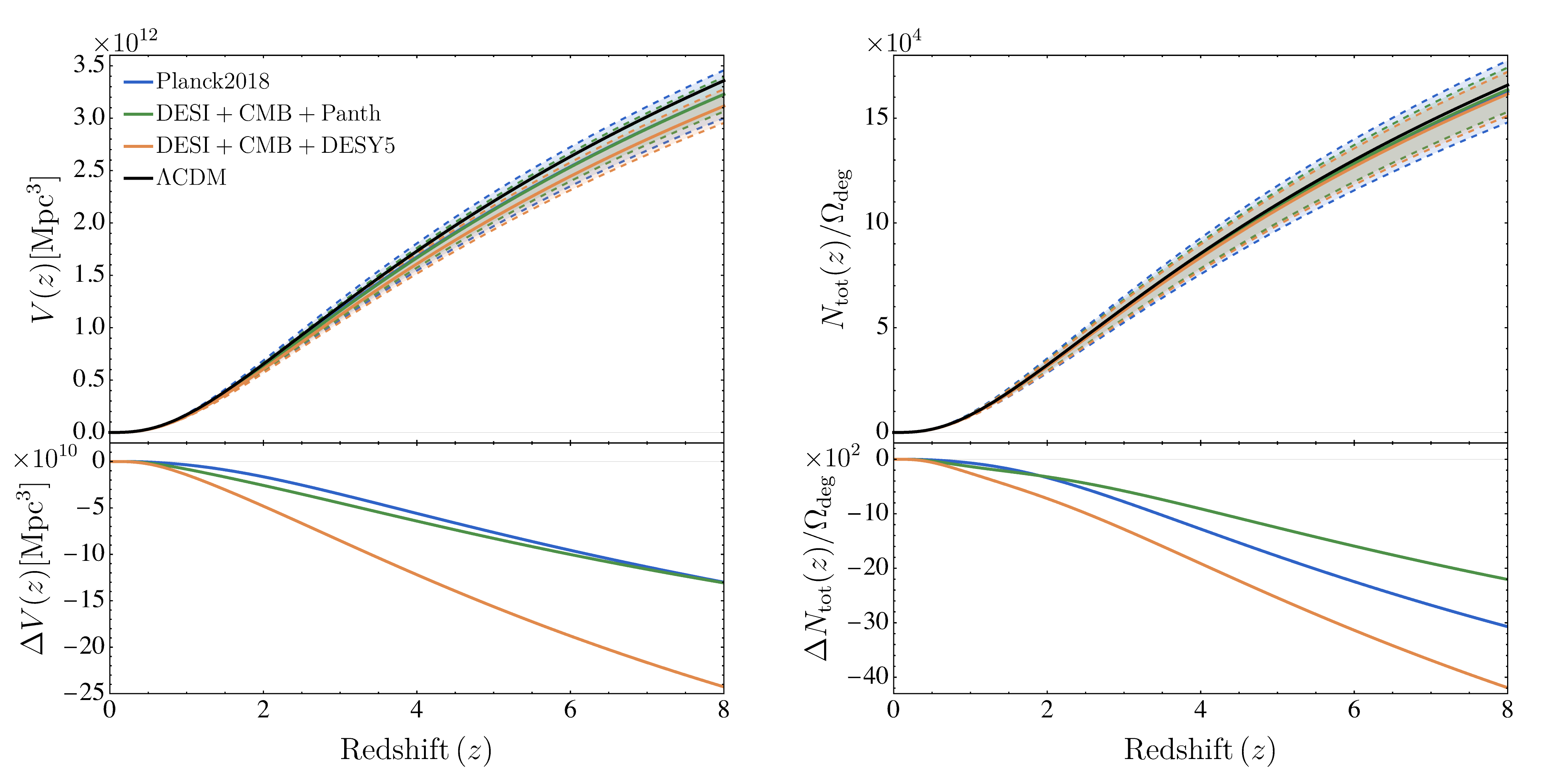}\vspace{-8mm}
    \caption{Total comoving volume (left, as given in Eq.~\eqref{eq:dv}) and number galaxy counts (right, as given by dividing Eq.~\eqref{Ntot1} by $\Omega_{deg}$) for the different parametrisations outlined in Table~\ref{tab:gwmodels}. Upper plots show the absolute values, taken to best fit our Universe (dashed lines corresponding to the errors as given in the corresponding references), hence explaining why they agree with $\Lambda$CDM results (black dashed line). Lower plots show the relative difference of each Quintessence best-fit parametrisation compared to $\Lambda$CDM. }
    \label{fig: V and Ntot}
\end{figure*}

\subsection{Galaxy Stellar Mass Distributions}

There are many ways in which this problem can be explored. In this paper we take a largely theoretical approach informed by empirical observations of galaxies to determine how and if this method can be used to derive cosmological parameters.  In this case we focus on an initial set of galaxies within a volume and then determine how the total number of galaxies per unit volume for these galaxies would change with time under various cosmological scenarios.

The first thing we consider in this process is the galaxy stellar mass function (SMF) which is defined as the number density of galaxies $\phi(M,z)$ in bins of stellar mass $\Delta M$ at each redshift ($z$).   This is related to the more traditional galaxy luminosity function which has long been a way to parameterise the distribution of galaxy luminosity \citep[e.g.,][]{Mortlock2015}.  The analytical SMF is determined observationally by counting galaxies at some redshift bin, evaluating their number density by dividing the counts by the comoving volume, and fitting the number density distribution as a function of mass with the Schechter function.

We use the stellar mass function as a basis for how many galaxies of different masses we have at an initial redshift, which can be an initial one at high redshift or a final one.  Practically, this initial redshift could be either at z = 0 and then evolved backwards in time, or at higher redshifts and forward evolved.  We use the stellar mass function as the basis from which we start our examination of how galaxy number and number density evolve when including additional astrophysics to galaxy formation. That is we use an initial observed fiducial mass function which we evolve at the high mass end due to known galaxy astrophysics building up stellar mass.  

We examine this as a way to quantify the number of galaxies at a given redshift and then evolve this number of galaxies through time with mergers and star formation which adds stellar mass to galaxies.

The integral over the stellar mass $M_*$ is the galaxy total number density $\phi_T(z)$, which gives a value for the cumulative number density of galaxies between masses $M_{\rm min}$ and $M_{\rm max}$ at a given epoch,  
\begin{equation} 
\label{phiT}
\phi_{\rm T} (z) = \int_{M_{\rm{min}}}^{M_{\rm{max}}} \phi (M, z)\, {\rm d}M.
\end{equation}
\noindent This integral is over the mass, $M$, which can be in a stellar format ($M_{*}$) or other forms of mass such as the total mass.  In this paper we use the stellar mass as a reference for the masses of galaxies. The stellar mass function is generally described empirically by the parametric formulation first introduced by \cite{schechter1976analytic} when describing our local universe. Since then the Schechter function has been adopted by most studies of galaxy evolution and mass assembly. It is usually given for galaxies in the nearby universe in a double form \citep[e.g.,][]{Baldy2012} such as: 
\begin{equation}
\phi(M,z) = \frac{e^{-M/M^{*}}}{M^*}\left[\phi_1^* \left(\frac{M}{M^*}\right)^{\alpha_1}+\phi_2^* \left(\frac{M}{M^*}\right)^{\alpha_2}\right].
\label{eq:schechter1}
\end{equation}



\noindent Eq.~(\ref{eq:schechter1}) describes two power laws of slope $\alpha_1$ and $\alpha_2$ at masses smaller than a characteristic stellar mass $M^{*}$ (given in terms of solar masses $M_\odot$). Beyond that scale the function is cut off by a high-mass exponential tail. The overall normalization is set by $\phi_1^{*}$ and $\phi_2^*$, which correspond to the number density at $M^{*}$ {and are $z$ dependent}. 

Typical values from field galaxies, {(i.e. at $z=0$),} are $M^{*}=10^{10.66}M_\odot$, $\phi^*_1=1.15 \times 10^{-2} h^3 \text{Mpc}^{-3}$, $\alpha_1=-0.35$, $\phi^*_2=2.3 \times 10^{-3} h^3 \text{Mpc}^{-3}$ and $\alpha_2=-1.47$ \citep[][]{Baldy2012}.  Just as for the number of galaxies, the value of the Schechter function parameters change with redshift due solely, it is assumed, to galaxy evolution processes, including star formation and merging. Practically $\phi(M,z)$ is found to evolve smoothly with redshift \citep[e.g.,][]{Duncan2014, Adams2023}, and as such most applications fit for each parameter independently at each redshift, hence they are normally treated as being redshift independent. 

It is usually assumed that the evolution of these parameters is due to galaxy evolution, with the cosmology  we used to calculate galaxy luminosities and masses being fixed and presumed correct.      Thus, it is in principle possible to use the evolution of the stellar mass function to infer how galaxy formation and evolution is occurring.  However, usually we do not use the method of mass or luminosity function changes to trace these formation processes, but instead, we can  measure galaxy star formation and merging through various direct methods as opposed to changes in the stellar masses of galaxies \citep[e.g.,][]{Madau1998, Conselice2003, Duan2024, Adams2023}. 

In this paper, we provide a new approach in determining this evolution, namely we account for the changes in the stellar masses of galaxies through star formation and galaxy merging to infer how changes in galaxy number density which differ from this expectation might be due to cosmological effects.  It is important to emphasis that this distinction  cannot easily be made with data  available today.  Later in this paper we consider how will we know the star  formation and history of mergers at higher accuracy with future surveys.  





\subsection{Galaxy Number Counts}

Eq.~(\ref{phiT}) enables us to calculate the total number of galaxies between two mass limits, at a redshift $z$.  Values typically used to cover all known galaxy types range from $M_{\rm{min}} = 10^{5}$ \msol to  $M_{\rm{max}} = 10^{12}$ \msol with redshift ranges from $z_{\rm{min}} = 0$ to $z_{\rm{max}} = 10$, the highest redshift where  thanks to JWST we will soon have statistically significant samples of galaxies with which to do tests such as this \citep[e.g.,][]{Harvey2024}. In the rest of this paper, since we are mainly interested in the case of massive galaxies, we will assume $M_{\rm min}=10^{10.5}M_\odot$. We use this limit as it ensures that we can measure the galaxy population down to this mass with a complete sample. It is also not too high such that we have enough galaxies to construct a large enough sample for which Poisson errors would not dominate.  With Euclid and Roman we will have even larger numbers of galaxies leading to even lower errors on the statistics than with JWST. 

Having obtained the galaxy counts for a fixed redshift, the next step is to find the total number of galaxies within a comoving volume  $V_c$, which is an integral over the number density distribution
$\phi_{T} (z)$ and the comoving volume of space, along with the redshift range we are interested in:
\begin{equation}
\label{Ntot1}
N_{\rm tot}(z) = \int^z_0 \phi_{T} (\hat{z})\, dV_c(\hat{z}) = \int^z_0 \int_{M_{\rm{min}}}^{M_{\rm{max}}} \phi (M, \hat{z})\, {\rm d}M \,dV_c(\hat{z})
\end{equation}
where the volume is integrated over the entire sky through
$4\pi$ steradians, and in this case from redshift $0$ to $z$.  Also we effectively integrate over the stellar mass here, although other forms of mass are also possible in principle.  




The comoving volume element of space is given by,
\begin{equation}
dV_c(z) = d_H \frac{d_T^2(z)}{H(z)/H_0} d\Omega dz,
\label{eq:dv}
\end{equation}
where $H(z)$ is the Hubble parameter with $H_0 = 100h$ km s$^{-1}$ Mpc$^{-1}$ the present day Hubble constant. $d_H=c/H_{0}$ is the Hubble distance, and 
\begin{equation}
  d_T(z)=c \int_0^z \frac{d\hat{z}}{H(\hat{z})}  \label{eq:comoving-dist}
\end{equation} 
is the proper (or co-moving) distance. In general we take $h=0.68$ from (\cite{Planck:2018vyg}), although as can be seen from Eqs.~(\ref{eq:schechter1}) and (\ref{eq:dv}) when substituted into Eq.~(\ref{Ntot1}) the $h$ dependence encoded in $\phi^*$ and $dV_c$ drops out at leading order for all cosmologies. {However it is clear from Eqs.~(\ref{eq:dv}) and (\ref{eq:comoving-dist}) that the comoving volume element depends on the cosmology, through the evolution of $H(z)$.} The comoving volume describes a volume of space in which the number of objects in the Hubble flow remains the same with redshift. 

By splitting this into shells of thickness $\delta z$, we can discuss the number densities in much smaller redshift bins. Hence we introduce shells, of thickness $\delta z$. The volume {between a redshift $z$ and $z + \delta z$} is simply  
\begin{equation}
V_{shell}(z) = \int_{0}^{z+\delta z} dV_c(\hat{z}) - \int_{0}^{z} dV_c(\hat{z}),
\label{eq:shellv} 
\end{equation}

\noindent In a similar manner to Eq.~(\ref{eq:shellv}), it is possible to derive the number of observed galaxies within shells of thickness $\delta z$:
\begin{equation}
N(z) = N_{tot}(z+\delta z)-N_{tot}(z).
\label{eq:Nfunction}
\end{equation}





Galaxy evolution affects $N(z)$ 
in three ways: galaxies with masses below $M_{\rm{min}}$ merge together, creating a galaxy with mass greater than $M_{\rm{min}}$, which increases the overall number; galaxies with mass above $M_{\rm{min}}$ merge together, which decreases their overall number; galaxies form stars, which increases their luminous mass, and results in some galaxies growing more massive than the lower mass limit in $\Phi_{\rm T}(z)$ of Eq.~\eqref{phiT}, leading to the increase in $N(z)$. 
All of these effects are described in greater detail in Section~\ref{galaxy-star}. 

\subsection{Cosmological Implications of Number Counts}
Before considering astrophysical effects on the galaxy counts there is another effect we need to discuss concerning the standard derivation of the luminosity function and the Schechter formalism which is intended to parameterise it. 
First it is important to recognise that the luminosity distances  for typical mass functions are typically obtained under the assumption of $\Lambda CDM$ cosmology. Given, that we are looking for evidence of cosmologies beyond the usual $\Lambda CDM$ framework, such as the impact dynamical dark energy may have on these functions, 
we need to be able to take this into account. We do this
via the following replacement in Eq. (\ref{phiT}), $M_{min} \to M_{min}^{'}$ where:
\begin{equation}
M_{min}^{'} = \frac{M_{min}}{q}.
\label{eq:modifymlow}
\end{equation}

\noindent where  $q \equiv (d_{L,\Lambda CDM}^2 / d_L^2)$, $d_{L,\Lambda CDM}$ is the luminosity distance in the standard $\Lambda CDM$ cosmology and $d_L$ is the same quantity in the new cosmology driven say by dynamical dark energy. For spatially flat universes ($\Omega_k=0$), $d_{L}\equiv d_{T}$ from Eq.~\eqref{eq:comoving-dist}.  
We are effectively introducing a new redshift dependent term into $\phi_T(z)$ through the revised minimum galaxy mass. We have to consider these limits when examining different cosmologies as we are aiming to match the same mass limits at different redshifts.  However, the mass measurements from observations typically assume $\Lambda$CDM, so these need to be accounted for through the variation based on the different luminosity distances within different cosmologies.

Our goal is to use this formalism to examine how the number of galaxies determined by Eq.~(\ref{eq:Nfunction}) {(or equivalently the total number in a given redshift range, Eq.~(\ref{Ntot1}))}, changes as a function of redshift for different cosmologies. The key change in the cosmological evolution comes through the Friedmann equation and the way this affects the luminosity distance calculations, hence the comoving volume element Eq.~(\ref{eq:dv}). We begin by writing the Friedmann equation as 
\begin{equation}
    H^2(z) = H_0^2 [ \Omega_{\rm{m}}^{(0)} (1+z)^3 + \Omega_{\rm DE}(z) ]
    \label{eq:Friedmann}
\end{equation}
where $\Omega_{\rm{m} }^{(0)}$ is the matter density parameter today, and $\Omega_{\rm DE}(z)$ is the density parameter associated with the dark energy. Assuming the dark energy is a perfect fluid with an equation of state $w(z)$, then for a flat cosmology containing only pressure-less matter and dark energy we can write 
\begin{equation}
    \Omega_{\rm DE}(z) = \left(\frac{H_0}{H(z)}\right)^2 (1-\Omega_{{\rm{m}}}^{(0)})  \exp{\left[3\int_0^z \frac{d\hat{z}}{1+\hat{z}} (1+w_{\rm DE}(\hat{z}))\right]}
\label{eq:eos}
\end{equation}
\begin{figure}
    \centering
    \vspace{-4mm}
    \includegraphics[width=1.05\linewidth]{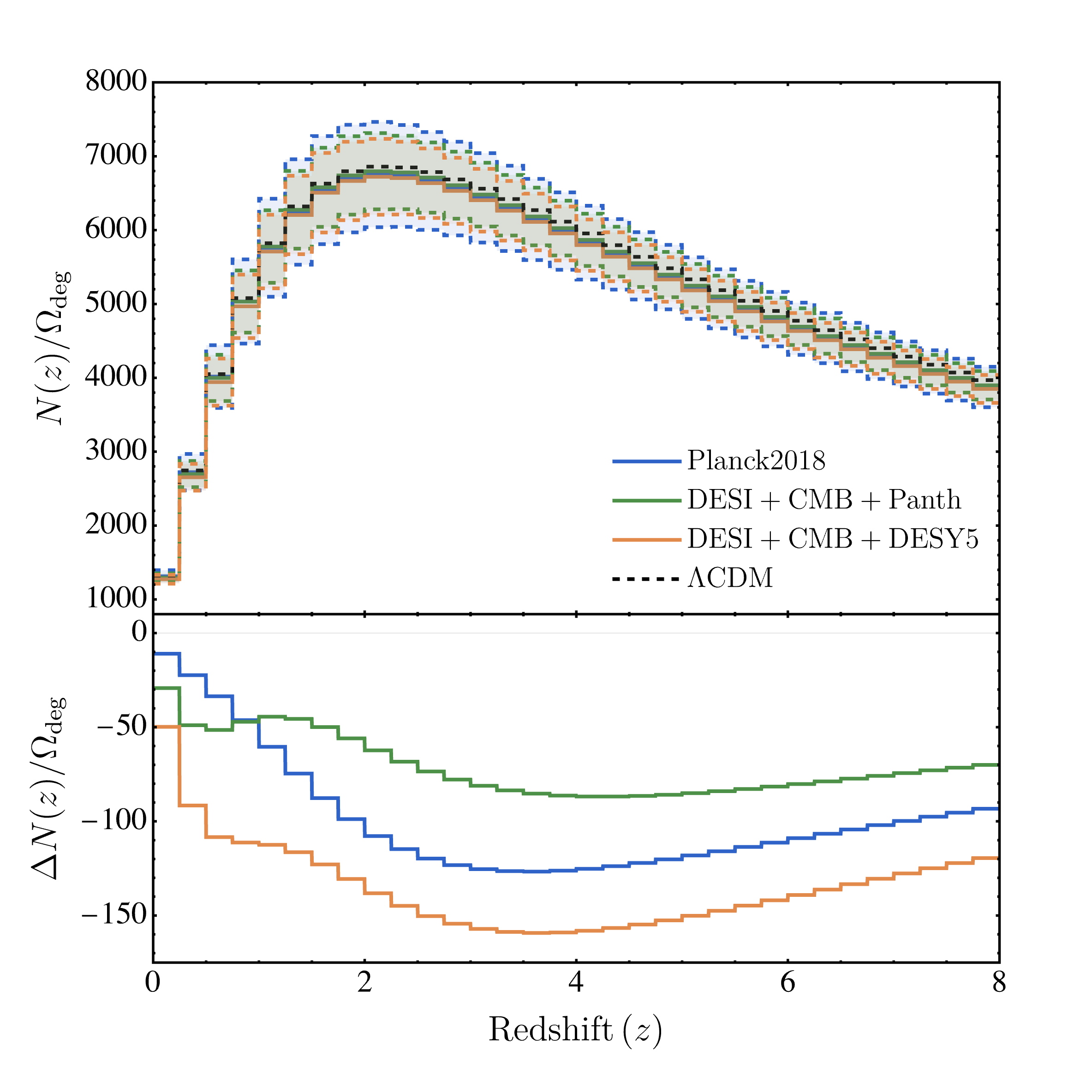} \vspace{-6mm}
    \caption{Number of galaxies per square degree within redshift bins of width $\delta z=0.25$, as given by Eq.~\eqref{eq:Nfunction}. On the top plot, the absolute values are shown, with $\Lambda$CDM corresponding to the black line and the colored lines to the best-fit parametrisations and errors from the models in Table~\ref{tab:gwmodels}. The lower plot shows the relative difference of each model to $\Lambda$CDM. }
    \label{fig: N Histogram}
\end{figure}
\noindent As a proof of principle for the technique, we will compare the galaxy number counts for a $\Lambda$CDM scenario, with a popular parameterisation of the equation of state for dynamical dark energy, namely the one proposed in \protect\cite{Linder_2003}  
\begin{equation}
w_{\rm DE}(z) = w_0 + w_a \frac{z}{(1+z)}
\label{eq:eos-gen}
\end{equation}
\noindent where $w_0$ and $w_a$ are constants. We note that this is one of the many parametrizations for $w_{\rm DE}$ in the literature, but is being adopted here mainly because it has become the go to parameterization  (for a review of the various models see \cite{Copeland:2006wr}).
This parameterisation has been used by both the recent Planck \citep[][]{Planck:2018vyg} and DESI \citep[][]{DESI:2024mwx} collaborations, and in what follows we will use their best fit values to see how they impact the predictions.

We now want to determine the galaxy number counts for models of dynamical dark energy parameterised by Eq.~\eqref{eq:eos-gen}, and by doing so to demonstrate the potential for this observable to constrain allowed models. In particular, we will compare a $\Lambda$CDM model with the best fit values for $w_0$ and $w_a$ coming from three experiments. The first is based on the Planck values obtained in \protect\cite{Planck:2018vyg}, and the other two are from the DESI combinations of DESI+CMB+Pantheon and DESI+CMB+DES Year 5 observations in \protect\cite{DESI:2024mwx}.

\begin{table}
\centering
\boxed{\begin{tabular}{|c|c|c|c|}

\textbf{Model} & \textbf{$\mathbf{w_0}$} & \textbf{$\mathbf{w_a}$}&$\Omega_{m}^{(0)}$ \\
\hline
$\Lambda$CDM  & $-1$ & $-$ & $0.3$\\
\hline
Planck'18  &  $-0.96 \pm 0.08$ & $-0.29 \pm 0.29$ & $0.315\pm0.007$\\
\hline
DCPa  & $-0.838 \pm 0.055$ & $-0.62 \pm 0.21$ & $0.3114\pm0.0057$\\
\hline
DCDY5 &  $-0.752 \pm 0.057$ & $-0.86 \pm 0.21$ & $0.3191\pm0.0056$\\
 
\end{tabular}}
\caption{The best fit values for $w_0$ and $w_a$ as given in Eq.~\eqref{eq:eos-gen}, as well as for $\Omega_{\rm{m}}^{(0)}$, for four different sets of data. The first row corresponds to a standard $\Lambda$CDM scenario with $\Omega_{\rm{m}}^{(0)}=0.3$, where as the other rows correspond to the 
Planck values obtained from \protect\cite{Planck:2018vyg} (second row), and DESI values from the combination of DESI+CMB+Pantheon observations (third row) and DESI+CMB+DES Year 5 in  \protect\cite{DESI:2024mwx} (fourth row). }
\label{tab:gwmodels}
\end{table}




\noindent The list of the relevant cosmological parameters can be found in Table~\ref{tab:gwmodels} including their best fit values for $\Omega_{\rm{m} }^{(0)}$.  

Our approach is straightforward. Using $w_{\rm DE}(z)$ from Eq.~(\ref{eq:eos-gen}) and suitably chosen parameters $w_0$, $w_a$ and $\Omega_{m}^{(0)}$ from Table~\ref{tab:gwmodels}, we solve for $\Omega_{\rm DE}(z)$ and $H(z)$ in Eqs.~\eqref{eq:Friedmann} and \eqref{eq:eos}, hence for the co-moving distance $d_T(z)$ in Eq.~(\ref{eq:comoving-dist}), the co-moving volume element $dV_c(z)$ in Eq.~(\ref{eq:dv}) and finally the number of galaxies within a sphere of radius $z$, $N_{tot}(z)$, in Eq.~(\ref{Ntot1}).

Figure~\ref{fig: V and Ntot} shows these quantities for the different  $w_0$, $w_a$ and $\Omega_{m}^{(0)}$ values shown in Table~\ref{tab:gwmodels}. Although $N_{\rm tot}(z)$ is an observable itself, in Figure~\ref{fig: N Histogram} we also present the number counts of galaxies within shells of thickness $\delta z=0.25$, per square degree, given by Eq.~\eqref{eq:Nfunction}, with $\Omega_{deg}=4\pi\times (180/\pi)^2$ being the solid angle of the entire sky in square degrees.  We use this thickness as it is a redshift range that is large enough to include enough galaxies, yet is also small enough so that very little evolution could have occurred in the galaxies.

Figure~\ref{fig: N Histogram} shows that the different possible cosmologies in use today can be distinguished with differences of up to of order 100 galaxies per square degree.  This effect is the highest at $z \sim 2-4$, although this is a purely theoretical best case scenario which does not take into account uncertainties in number counts, or in the unknown effects from the astrophysics of galaxy evolution.
Although most of these functions closely agree with the predictions of $\Lambda$CDM, this is expected since the values in Table~\ref{tab:gwmodels} are selected to best describe our universe in terms of consistency with large survey observations of the CMB, galaxy clustering, and galaxy weak lensing. 

Having established the theory behind the number counts and how the differing cosmologies can in principle be constrained with them, we now move on to discuss two important astrophysical features which impact the galaxy number counts, namely galaxy mergers and star formation within a galaxy.

\section{The impact of galaxy mergers and star formation on galaxies number densities} \label{galaxy-star}

\subsection{Galaxy Evolution Effects} \label{subsec:gal-evol}
In determining the impact of star formation rates and galaxy mergers on the redshift dependence of the galaxy number counts we begin by making the following assumptions:
\begin{enumerate}
\item We assume that there are no new galaxies formed below the highest redshifts we examine. For practical purposes we assume that all galaxies exist in some form at $z_{\rm max} = 8$ and no `new' galaxies form and become massive over subsequent cosmic time. As we see many very massive galaxies at high redshift, it is unlikely that very massive galaxies are `born' at redshifts lower than this \cite[e.g.,][]{Mortlock2015, Harvey2024}.

\item We assume that galaxies grow in mass and do not lose mass with time.  The two major ways in which we consider how the stellar mass grows are through star formation and merging activity. Some basic ideas behind these processes are discussed in the appendix of \citet{Conselice2003}.   In terms of gas masses, star formation rates and accretion of gas, this formalism is used in detail in papers such as \citet{Conselice2013} and \citet{Ownsworth2016}.  This previous work used continuity equations similar to the ones here to determine how galaxy formation is occurring via gas accretion as well as how galaxies are assembled by various processes including major/minor mergers and gas accretion to establish the baryons for future star formation \citep[][]{Ownsworth2016}.


\end{enumerate}
\subsubsection{\label{subsec:sfr} Star Formation Rates and History}

The star formation rate (SFR), {$\psi(z)$}, determines the amount of mass produced as the result of star formation, and is obtained by observing light emitted by young stars or light emitted by the effects of young stars through dust emission and radio emission. 
 Traditionally, it is represented by using the function fitted to the entire star formation history as provided by \citet{Madau2014}, 

\begin{equation}
\psi(z)=0.015\,{(1+z)^{2.7}\over 1+[(1+z)/2.9]^{5.6}}\,
\label{eq:sfrd}
\end{equation}

\begin{figure}
    \centering
    \vspace{-4mm}
\includegraphics[width=0.99\linewidth]{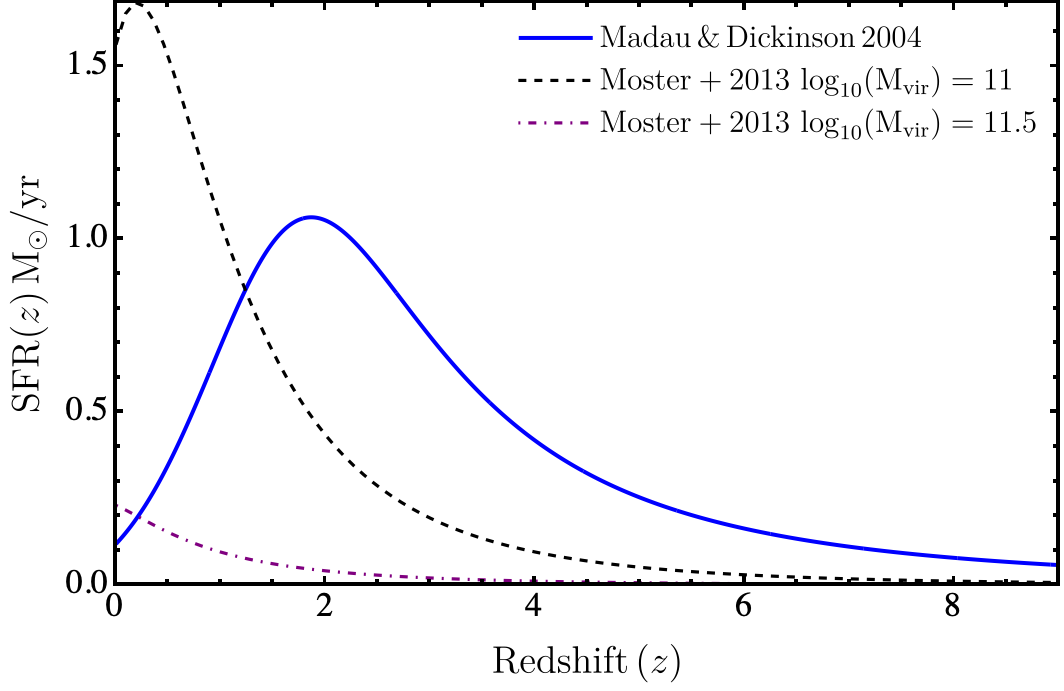} 
    \caption{Plot of the average star formation rate per galaxy which we use in this paper. We show a modified version of the standard Madau \& Dickinson star formation history (blue line) and the modified versions, based on theory (see text), designed such that the integral of the star formation provides a total stellar mass of M$_{*} = 10^{11}$ M$_{\odot}$ or M$_{*} = 10^{11.5}$ M$_{\odot}$ over redshifts from $0 < z < 10$.   These forms use the models of \protect\cite{Mostar2013}. Within this paper, we assume M$_{*} = 10^{11}$ M$_{\odot}$ is the typical final stellar mass of a galaxy in our sample.}
    
    \label{fig: SFR}
\end{figure}

\noindent where this star formation history is fitted to the observable data of an integration of star formation rates over all masses at different redshifts. The standard units in which SFR is measured are solar masses per year.   We show this fit as a blue line in Figure~\ref{fig: SFR}.  However, note that in \citet{Madau2014}, the star formation rate density is measured over galaxies at all masses, whereas we are interested in the more massive galaxies and individual galaxy star formation histories.  

It turns out that very few galaxies have individual star formation histories that follow this functional form.  It is also important to note that galaxy star formation histories are complex and in general cannot be accounted for simply by analytic functions. Fortunately, on average, or if considering an ensemble of galaxies, it is actually possible to write a functional form such that the average star formation rate can be parameterised. 

The star formation history of galaxies as a function of the stellar mass is not well characterized as a function of redshift up to $z \sim 8$ in terms of empirical observations.  This is due to the fact that the star formation rates that we can measure at high redshift are typically only for the brightest galaxies, and we are incomplete for the fainter systems that are not undergoing a high amount of star formation \citep[e.g.,][]{Whitney2020, Adams2023}.

Thus, we incorporate into our calculation, star formation rates that are taken from abundance matching from \cite{Mostar2013}.  
 We use the fitted formula in this paper to represent a typical SFR history for  the more massive  galaxies ($\psi_{\rm{MG}}(z)$), namely we use: 


\begin{eqnarray}
\psi_{\rm MG}(z) &=&\frac{{\rm d}M_*}{{\rm d}t}(z, M_{\rm vir}) = \nonumber \\  
&\hspace{-.6cm}=&\hspace{-.6cm}
 f_1(M_{\rm vir}) \, (1+z)^{-f_2(M_{\rm vir})} \,
\exp\!\left( \frac{f_2(M_{\rm vir})}{f_3(M_{\rm vir})} \cdot \frac{z}{1+z} \right)
\label{SFR-massive-gal}
\end{eqnarray}

\noindent where the following functions are used,

\begin{align}
f_1(M_{\rm vir}) &= f_{10} \, \exp\!\left[ -\frac{\big(\log_{10} M_{\rm vir} - f_{11}\big)^2}{2 f_{12}^2} \right], \\
f_2(M_{\rm vir}) &= f_{20} + f_{21} \, \log_{10}\!\left( \frac{M_{\rm vir}}{10^{12} M_\odot} \right), \\
f_3(M_{\rm vir}) &= 10^{\, f_{30} + f_{31} \left( \frac{M_{\rm vir}}{10^{12} M_\odot} \right)^{f_{32}} }.
\end{align}

\noindent The parameters used in the star formation rate model are:
\[
\begin{aligned}
& f_{10} = 2.658, \quad f_{11} = 11.937, \quad f_{12} = 0.424, \\
& f_{20} = 5.507, \quad f_{21} = 2.437, \\
& f_{30} = -0.915, \quad f_{31} = 0.696, \quad f_{32} = -0.159.
\end{aligned}
\]

\noindent We obtain the star formation rate, and how it evolves with redshift,  from these equations, using a fiducial example of a system with a virial mass of $M_{\rm vir}=10^{10.5}$ \msol, which we use throughout.  
Note also that whilst we examine features in terms of the stellar mass when integrating the mass function, the star formation history we use is parameterised by the virial mass.    The relation between the virial mass and stellar mass is not trivial to determine as a function of redshift, but has a relation of a factor of 5-10 \citep[e.g.,][]{Conselice2018}.


This form of the star formation rate with time is, we repeat, an average over all systems.  This is sufficient and even ideal for our purposes as we are interested in the effects of star formation rate over the entire galaxy population and using this function will allow this to be determined.   The parameterisation of Eq.\eqref{SFR-massive-gal}
is also useful as it gives us a low SFR at $z\sim 0$ consistent with the observations of massive galaxies. 


We later show below in \S 3.2 using  Eqs.~(\ref{deltaM-SFR}) and (\ref{eq:nmu1}), how the star formation rate $\psi_{\rm{MG}}(z)$ is used to determine the stellar mass formed over a given redshift period.   These values can then be used to adjust the volume densities of galaxies at a given redshift. 

A critical part of this analysis is the formation of stellar mass through star formation.  This is simply the integral of the star formation rate ($\psi$) over time.  The general equation that we use is:

$$ \delta M(z(t_1,t_2)) = \int_{t_1}^{t_{2}} \psi_{\rm {MG}} d{t},$$

\noindent where the integral over the star formation rate with time ($t$) represents the change in mass in a single galaxy. For this calculation we assume the star formation rate is known.  Observationally this is not always straightforward, as there are issues with the initial mass function and other stellar population features which can change the resulting conversion of observations to inferred star formation rate \citep[e.g.,][]{Harvey2024, Austin2026}.
This star formation integrated over time is also not technically the resulting stellar mass over a long period of time.  Some fraction of the stellar mass created will be returned to the interstellar medium (ISM) of the galaxy.  We include this fraction of the mass returned in our calculations of the amount of total resulting stellar mass retained in the galaxies \citep[e.g.,][]{Conselice2013, Harvey2024}.


\subsubsection{\label{subsec:merger} The Galaxy Major Merger Rate}

Galaxy mergers are now known to be a major way in which galaxies form even up to the highest redshifts \cite[e.g.,][]{Conselice2014morph, Mantha2018, duncan2019, Duan2024}.   These mergers can be divided into those between two galaxies with a similar mass, called major mergers, and galaxies which merge with those of lower mass called minor mergers. Many studies have investigated major mergers out to high redshift \citep[e.g.,][]{Conselice2003, Bluck2009,Bluck2012, Mundy2015, Mantha2018, duncan2019,Ferreira2020, husko2022, shibuya2022, Duan2024}. The same cannot be said yet of the {minor mergers with smaller host-to-companion stellar mass ratios.  These minor
mergers are typically defined as galaxy pairings of mass $M_1 < M_2$, satisfying $0.1 < \mu \equiv (M_1/M_2) < 0.25$}, with major mergers satisfying $0.25 < \mu <1.0$.
The deep imaging, increased sample sizes, and high completeness needed to study these minor mergers has made it such that only recently has it been possible to study their evolution and determine their role in galaxy formation \citep{Bluck2012, Man:2014jsa, Lofthouse2017, Shah:2020jmc}. 

Furthermore, mergers are predicted in most cosmological models to be a significant method of forming galaxies, possibly triggering AGN and central black hole formation, and for instigating star formation \citep[e.g.,][]{Kaviraj2015, Shah:2020jmc}. In addition, to fully understand galaxy formation, and test cosmological models in detail, the merger rate will need to be measured and fully accounted for. Given those uncertainties, in what follows we concentrate on the impact of major mergers on the galaxy number counts, and only consider minor mergers where they lead to the formation of massive galaxies satisfying $M>M'_{\rm min}$ where we recall $M'_{\rm min}$ is defined in Eq.~\eqref{eq:modifymlow}. 

To determine the rate of major mergers, i.e. mergers of massive galaxies of comparable size, we use an analytical representation that works well over a wide range of redshifts as described in \citet{Conselice2008} and most recently fit with JWST data in \citet{Duan2024}
\begin{equation}\label{LargezRofz}
R(z) = R_0(1+z)^m \times {\rm exp}(\gamma*(1+z))
\end{equation}
where $R(z)$ is the rate of mergers per galaxy per gigayear, $R_0$ is the merger rate today, with $m$ and $\gamma$ being constants.  Note that the merger rates we use are empirical and derived from observations.  The time-scales for these mergers are astrophysical and do not depend on the cosmology assumed as the merger process is driven by dark matter and internal dynamics \citep[e.g.,][]{Whitney2021, Duan2024}.  However, the amount of time between redshifts and the look-back time does depend on cosmology and thus this is included when we integrate the merger rate back through time.

Throughout this analysis, we investigate how the deviations in number counts depend on the error measurements of the various quantities that are needed to infer the evolution of galaxies through the star formation rate and merger history. To characterise this, we use a parameter $\varepsilon$, which allows us to change the value of the uncertainty on these parameter fits in the galaxy formation and evolution history.  A value of $\varepsilon = 1$ implies that we are using the latest uncertainties on the values of the fitted parameters.  We use lower values of $\varepsilon$ to determine how features would change when we have a lower uncertainty in the value of the measurements of the parameters in these fits.  This will naturally occur once these parameters are measured using future and ongoing surveys such as Euclid, LSST and Roman and when better models of star formation and merging timescales are available.  

Merger fractions are now measured up to $z=12$ \citep[e.g.,][]{Duan2024}, and in general in these observations, we find that it increases up to high redshifts, but later becomes more or less constant for $z > 2$. For this work, in Eq.~\eqref{LargezRofz} we use the values determined in
\citet{Duan2024}, namely $R_{0} = 0.013\pm  0.007{\varepsilon}$,
$m = 3.36\pm 0.73{\varepsilon}$ and $\gamma = -0.14\pm 0.18{\varepsilon}$.  
As we show below, using these values to probe the evolution of the merger rate has a profound effect on the number of galaxies at a given redshift, hence on the number counts. 


This effect is twofold. The most obvious one is that mergers decrease the visible number of galaxies in the sky. However, a galaxy that forms due to a merger is more massive than galaxies that participated in the merger. This could lead to the second effect where two $M < M_{\rm min}'$ galaxies, not accounted for before, merge to form an $M > M_{\rm min}'$ galaxy,
which would imply an increase in $N(z)$. 
To take these two effects into account, we utilise the findings of \cite{Mundy:2015qgi}: who find the mass gain from a major merger to be 25\% on average, which is consistent with examining the mass accretion from a secondary galaxy during a galaxy merger.


\subsection{\label{subsec:together} Combining star formation and galaxy merger rates}

In this section we investigate how {the combination of} the star formation rate and galaxy merger rate change the number of galaxies in a given population. To proceed, we divide out by the volume in which we examine this evolution, such that we consider the number density of galaxies at a specific redshift. This quantity is given by the integrated Schechter function in Eq.~\eqref{eq:Nfunction}, which in its simplest form is constant in the absence of galaxy evolution. However, in this section, we will add the corrections that will introduce redshift dependence.

 We begin by exploring the role of star formation on the observed number density, assuming that galaxies of different masses have the same average star formation history within a large enough time-span. We recall that the star formation rate at a given redshift ($z$) is 
 given by Eq.~\eqref{SFR-massive-gal}. Given that star formation increases the luminosity and stellar masses for all galaxies, it is effectively equivalent to increasing the average masses of galaxies by an amount $\delta M$  between $z_{\rm max}$ and $z$, where $z$ is the redshift of observation: 
 
\begin{equation} \label{deltaM-SFR}
\delta M(z) = \int_{z}^{z_{max}} \psi_{\rm {MG}}(\hat{z}) \frac{1}{(1+\hat{z})H(\hat{z})}d\hat{z}, 
\end{equation}
which of course implies $\delta M(z_{\rm max})=0$. In our formulation the stellar mass of galaxies are all {\it increased} by an amount $\delta M$.  This is certainly not precisely correct, given the variable star formation histories of high redshift galaxies \citep[e.g.,][]{Harvey2024}, but it is an approximation that is, on average, correct. In this situation therefore the lower limit of the integrand for the number density of galaxies changes {to} $(M'_{min} - \delta M)$, {as opposed to $M'_{min}$ as in section~\ref{sec:theory}}. This is because galaxies that were once too low mass to be counted are now above the threshold with this mass difference given by Eq.~\eqref{deltaM-SFR}. 
Hence the additional number of galaxies formed due to this star formation process between any two redshifts $z_i$ and $z_f$ is given by:

\begin{equation} 
\delta n_{sf}(z_i,z_f) = \int_{M'_{min}(z_i) - \delta M(z_i)}^{M'_{min}(z_{f}) - \delta M(z_{f})} \phi(M) dM .\label{eq:nmu1}
\end{equation}
where recall $\phi(M)$ is given by Eq.~\eqref{eq:schechter1}. Note that usually $\delta M(z_f) = 0$, whereby no mass has accumulated on average as we are considering this as the beginning of the redshift bins. 


\noindent We next consider the modification of the number of galaxies due to galaxy mergers. First, we consider the number of galaxies decreasing between our mass limits of $M'_{min}(z)$ and $M'_{max}(z)$. The way to calculate this is through the merger rate $R(z)$ in Eq.~(\ref{LargezRofz}). The value of $R$ can be thought of as the fraction of galaxies within a given time scale or redshift range that have merged. This will reduce the number of galaxies by the integral of $R(z)/2$.  This is because every major merger reduces two galaxies to one, where multiple mergers can be considered rapid subsequent single mergers.  With this understanding, the decrease in the number of galaxies due to mergers between redshifts $z_i$ to $z_f$ where $z_i<z_f$ is:

\begin{equation} \label{eq:nmu}
\delta n_{m,u}(z_i,z_f) = - \frac{1}{2}\phi_T(z_f)  \int_{z_i}^{z_f} R({z}) \frac{1}{(1+{z})H({z})}d{z}, 
\end{equation}
\noindent where the value of $\phi_T(z_f)$ corresponds to the number density of galaxies at redshift $z_f$ and is given by Eq.~(\ref{phiT}).  Note that this value is also negative, meaning that it reduces the number of galaxies within a mass range.  Here we make the assumption that the merger rate does not depend on the stellar or total mass of galaxies.  This is generally not true over large ranges of stellar masses \citep[e.g.,][]{Casteels2014}.  However, within the high mass galaxy range we consider here, this approximation is appropriate to this new cosmological parameter measurement. 

We also have to consider the number of mergers that occur below the nominal limit of $M'_{ min}$. To do this, we assume that the average increase in mass in a merger is given by $\eta <1$, such that after a merger, galaxies of mass $M$ will increase their mass by the amount $\eta M$.  Thus, a merging galaxy with a mass below the lower limit by the amount $M'_{min}(1 - \eta)$ will enter the required range of masses after the merger. 

Since the integral of $R$ gives the fraction of galaxies that undergo a merger, the increase in the number of galaxies due to these mergers is given by: 
\begin{equation}\label{eq: nml}
\delta n_{m,l}(z_i,z_f) = \int_{z_i}^{z_f} \int_{(1 - \eta) M'_{\rm min}(z)}^{M'_{\rm min}(z)} \frac{R(z) \phi(M,z)}{(1+{z})H({z})} dM dz
\end{equation}

\noindent where the value of $(1- \eta)$  times the lower mass limit gives a measure of how many galaxies that were too low mass in earlier epochs would now be in our lower mass limited sample after they have merged and grown in mass.  {Note that in Eq.~\eqref{eq: nml}, because of the non-trivial $M$ dependence in the integrals, we use $\phi(M,z)$ instead of $\phi_T(z_f)$, which does not contain the information on galaxy mergers and star formation. 
This approximation is valid as long as $\phi_T(z)$ does not become small enough for low–mass galaxies to merge more frequently than high–mass galaxies. For the models considered here, this condition is satisfied.} 

{Combining the results from star formation rates and the two elements of galaxy mergers we see that} the total number of galaxies at a redshift lower than $z_f$ can be written as:
\begin{equation}
\begin{split}\label{eq: phiT(zi)}
\phi_T(z < z_f) &= \phi_T(z_f) + \delta n_{sf}(z,z_f) \\&+ \delta n_{m,u}(z,z_f) + \delta n_{m,l}(z,z_f).
\end{split}
\end{equation}

\noindent The number of mergers in a given period of time depends on the density of galaxies, which is described by $\phi_T$. In the absence of mergers and star formation this quantity is constant for $\Lambda$CDM.  However, this is not the case for more generic quintessence scenarios, like the ones we described in \S~\ref{sec:theory}. In either of those scenarios, star formation and galaxy mergers will introduce an effective time dependence on $\phi_T$, affecting the evolution of the density of galaxy counts. Therefore, to consistently infer the number of mergers at a given redshift, we calculate the corrections in $\phi_T$ (now $\phi_T(z)$) iteratively by taking small steps of the redshift $\delta z$. 
In particular, to calculate the evolution from $z_{\rm max}$ to $z=0$, we must calculate $\phi_T(z)$ at each step and use it to calculate the next step, $\phi_T(z-\delta z)$.

\begin{figure}
    \centering
    \includegraphics[width=0.99\linewidth]{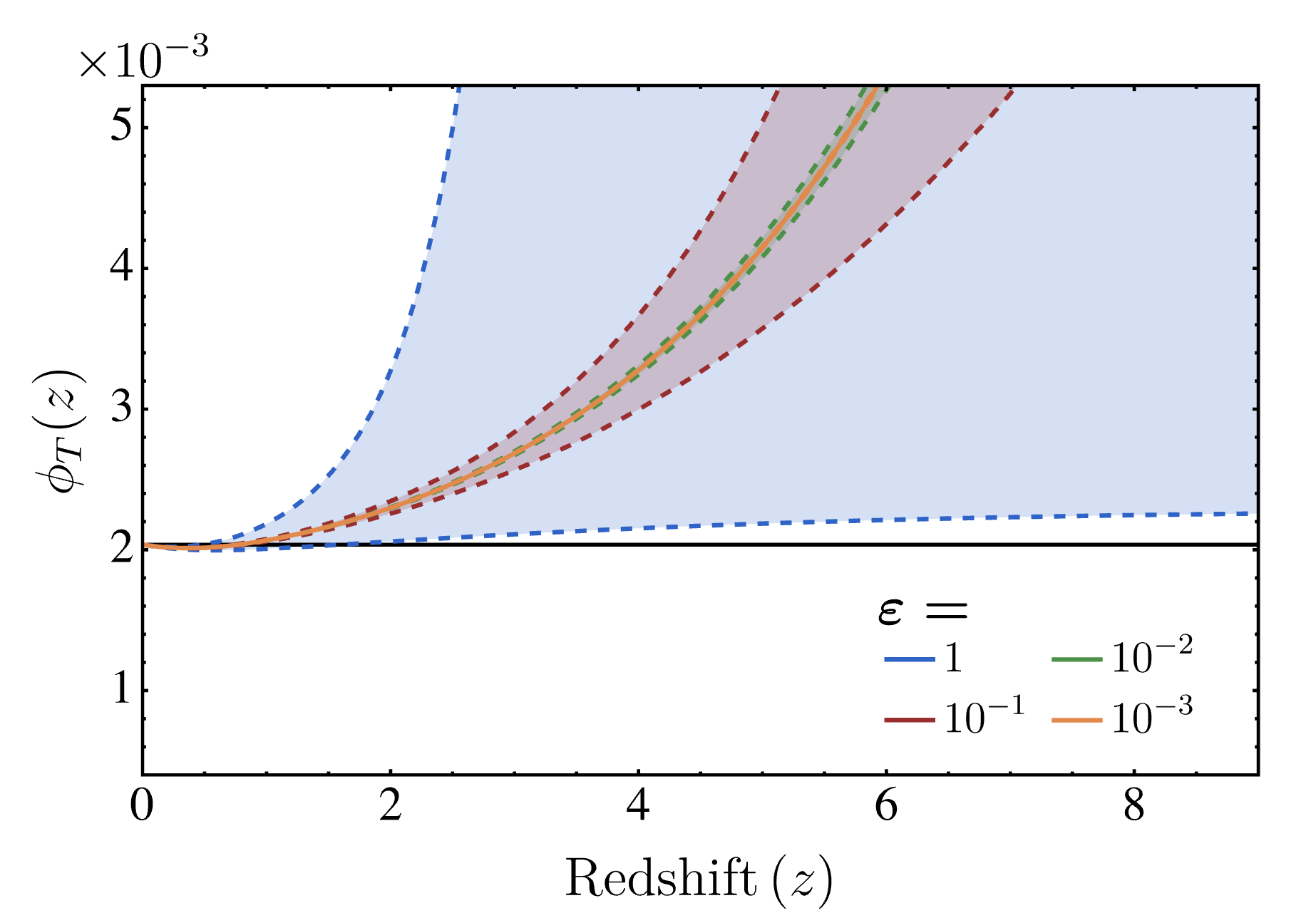}
    \caption{Evolution of the total galaxy number density as given in Eq.~(\ref{eq: phiT(zf)}) for a $\Lambda$CDM cosmology, including star formation effects and galaxy mergers, with step size $\delta z = 0.25$. The black line shows the constant Schechter function with no galaxy evolution, as calculated in \S~\ref{sec:theory}. Coloured regions show the effect that the errors on merger rates have on the evolution of the Schechter function, parametrised by $\varepsilon$ as shown in the text below.}
    \label{fig: phiT LCDM}
\end{figure}
Given an initial galaxy number density function $\phi_T(z_{max})$, we calculate its evolution until today. However, { in reality our best galaxy stellar mass function measurements are local, corresponding to $z \sim 0$, hence running the simulation backward in time is more effective in that we will be starting from more reliable initial conditions}. For that, solving for $\phi_T(z_f)$ in Eq.~\eqref{eq: phiT(zi)}, we obtain 
\begin{align}\label{eq: phiT(zf)}
\phi_T(z_f) =& \frac{\phi_T(z_i) - \delta n _{sf}(z_i,z_f)-\delta n _{m,l}(z_i,z_f)}{\left(1-\frac12\int_{z_i}^{z_f} R({z}) \frac{1}{(1+{z})H({z})}d{z}\right)},
\end{align}
where we have expanded $\delta n_{m,u}(z_i,z_f)$ following Eq.~\eqref{eq:nmu}.

\begin{figure}
    \centering
    \includegraphics[width=0.99\linewidth]{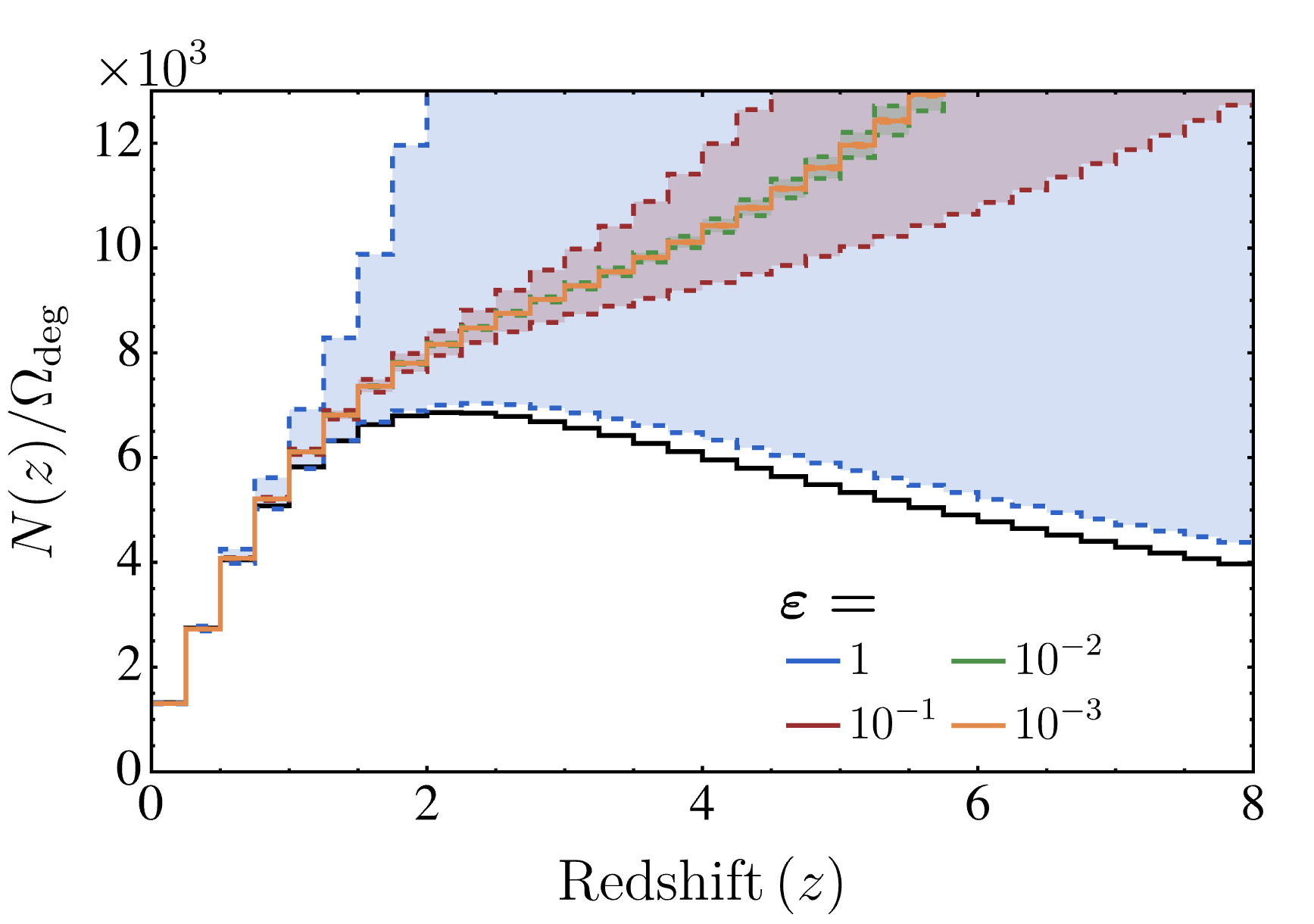}
    \caption{Number of galaxies per square degree within redshift bins of width $\delta z=0.25$, as given by Eqs.~\eqref{eq:Nfunction} and (\ref{eq: phiT(zf)}) for a $\Lambda$CDM cosmology. As with Figure~\ref{fig: phiT LCDM}, the black line shows the galaxy counts with no galaxy evolution history, as calculated in \S~\ref{sec:theory}. Coloured regions show the effect that the errors on merger rates have on the galaxy counts, parametrised by $\varepsilon$ as shown in the text below. }
    \label{fig: N Histo Mergers}
\end{figure}

As an example of the technique, in Figure~\ref{fig: phiT LCDM}, we have plotted Eq.~(\ref{eq: phiT(zf)}), the modified number density of galaxies, for a $\Lambda CDM$ cosmology, with $M_{\rm min}=10^{10.5}M_\odot$,  incorporating the errors presented in the text as limiting cases. While current errors dominate the results, we can see that if in the future we could lower them by a factor of $\sim$ 10-100, it would make galaxy number counts a very competitive method with which to constrain the expansion history of our universe. In Figure~\ref{fig: N Histo Mergers}, we show the equivalent calculation for the number of galaxies in redshift bins of $\delta z = 0.25$, given by $N(z)$, in Eq.~\eqref{eq:Nfunction}.

As can be seen in Figure~\ref{fig: phiT LCDM} and in Figure~\ref{fig: N Histo Mergers} this method can be used to determine the merger and star formation history. In the future this method will be useful to measure both the cosmology and the merger and star formation history, independent of direct measurements, by examining the change in number density with time for galaxies.   This however is years away and will require a more detailed knowledge of the number counts and how they vary across redshifts. 



\section{Results} \label{Sec:Results}

\subsection{Comparing $\Lambda$CDM and Quintessence with the modified $\phi_T$.} \label{Sec:compare-models}

Thus far, we have considered how different cosmologies impact on the number counts of galaxies through three key effects -- the change in the way the co-moving volume evolves with redshift, the impact of star formation on the galaxy counts and the impact of galaxy mergers. 

These have culminated in a modified integrated luminosity function Eq.~(\ref{eq: phiT(zf)}) which has directly incorporated the star formation as well as galaxy merger effects,  and the number counts $N(z)$ of Eq.~(\ref{eq:Nfunction}) which also includes the volume dependence as well as the other two contributions. We have seen that although the current levels of uncertainty in some of the parameters make it unrealistic to use number counts as a means of differentiating between cosmologies at the moment, in the future as these uncertainties reduce, it will become more competitive. 

With this in mind, we now move on to show a proof of principle for this claim, by comparing the results of $\Lambda$CDM with the most commonly used parameterisation of dynamical dark energy, Eq.~\eqref{eq:eos-gen}. We remind the reader 
that within that parameterisation, tentative evidence for phantom like behaviour has been suggested in the recent DESI data \citep[][]{DESI:2024mwx,DESI:2025zgx} through the  $w_a$ coefficient in Eq.~(\ref{eq:eos-gen}.  
In what follows, we take two of the DESI best fit values for the equation of state parameters, along with the Planck best fit value \cite{Planck:2018vyg} for the same parameterisation. 

We stress we are not trying to suggest these are the definitive values but want to compare their number count predictions against the $\Lambda$CDM scenario (which recall is $w_a=0$), in order to see whether galaxy counts including evolution will ever be able to tell state of the art cosmological calculations apart. A summary of the parameter values of the three models are given in Table~\ref{tab:gwmodels}.
For this comparison, we do not include the associated errors quoted for $w_0$ and $w_a$. This is because we are assuming we know the background cosmology and want to determine the impact of the errors associated with the numbers counts themselves through galaxy evolution. 

\begin{figure*}
    \centering
    \includegraphics[width=1\linewidth]{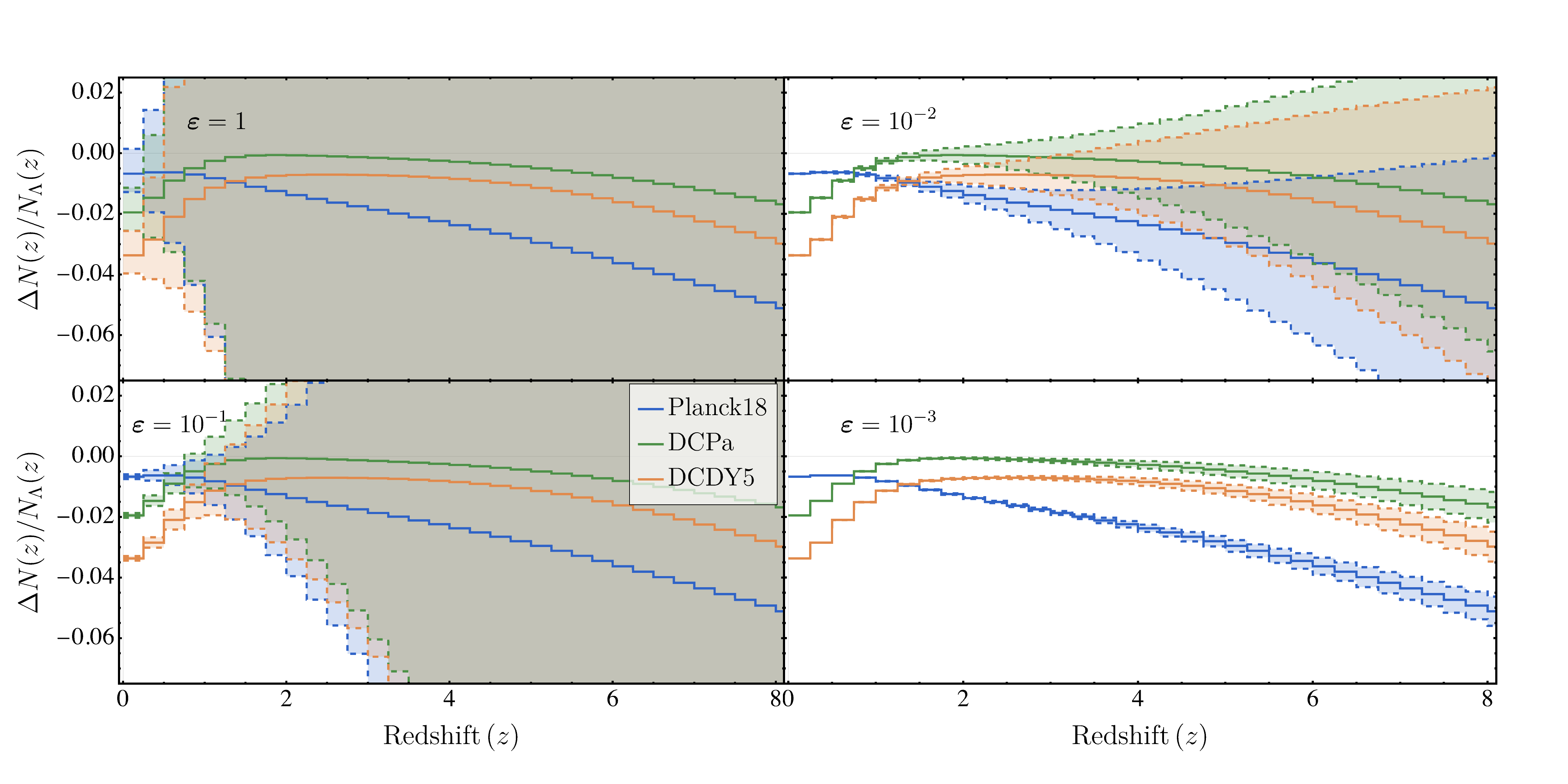}
    \caption{The ratio between the difference in the number counts in $\delta z=0.25$ bins of the three dynamical dark energy cosmologies in Table~\ref{tab:gwmodels} and $\Lambda$CDM, divided by the number counts $N_\Lambda (z)$ in $\Lambda$CDM as a function of redshift $z$. The four figures correspond to different values of the error parameter $\varepsilon$.  The lines in the plot show the values for the Planck18, DCPA, DCDY5 cosmologies.    Each panel shows the difference in the predicted volume for that cosmological model subtracted by the $\Lambda$CDM value, then normalised by the corresponding value in $\Lambda$CDM. 
    }
    \label{fig: N merger comparison}
\end{figure*}

Figure~\ref{fig: N merger comparison} directly compares the three models against $\Lambda$CDM. In particular we show the  difference between the number counts of each parameterisation compared to that of $\Lambda$CDM, divided by the number count from $\Lambda$CDM. The four enclosed figures simply correspond to four chosen values of $\varepsilon$. Of note in Figure~\ref{fig: N merger comparison} is the trend amongst the dynamical models, namely they all give a systematically lower count than that of $\Lambda$CDM with a difference of order 4\% at small redshift but changing to a minimum 1\% or so around $z \sim 1$, before rising again to a difference of order 5\% as $z \to 8$. It is interesting that the Planck 18 parameterisation which has the closest $w_0$ value to that of $\Lambda$CDM, and the least negative value of $w_a$, begins close to that of $\Lambda$CDM for small redshifts but then drifts further away from it compared to the DESI cases as $z$ increases. This small redshift evolution of $\Delta N/N_\Lambda$ could well be a significant probe for dynamical dark energy. However once again we caution the reader, as can be seen from the upper right most figure, to be able to tell apart the models, we require galaxy evolution errors to be much smaller than they are today.

To summarise our findings, it will be possible to determine the features of an evolving dark energy through examining galaxy counts within an evolving galaxy population.   Essentially what we have done is to remove the galaxy evolution component form the volume densities which reveals how a pure population of systems would change with volume only due to the effects of a different volume evolution provided by different dark energy parameterisations.   As can be seen however, this will however require accuracies in the galaxy formation history parameters to be known much more accurately than we do today.

We conclude this section with a comment on the role of the matter density $\Omega_m$. One interpretation of the driving force behind the quoted preference for dynamical dark energy in \cite{DESI:2024mwx} and \cite{DESI:2025zgx} is that their best fit $\Lambda$CDM models are in tension with the matter density from Planck \citep[][]{Planck:2018vyg}. It raises the interesting question, how do the number counts of galaxies depend on the matter density? This is trivial to see through the Friedmann equation Eq.~\eqref{eq:Friedmann} which in turn affects the distance estimates. 

\begin{figure}
    \centering
    \includegraphics[width=1\linewidth]{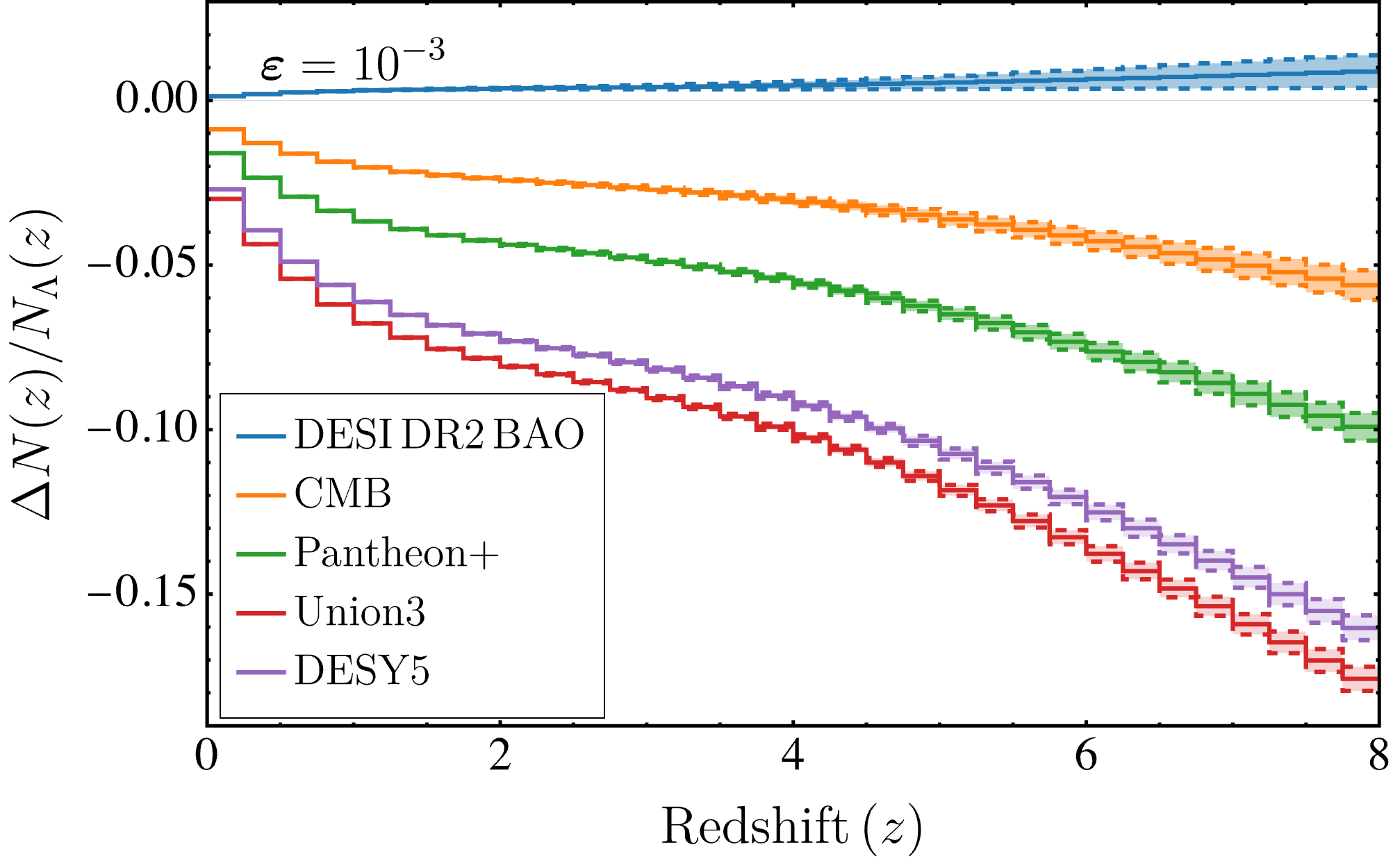}
    \caption{Effect of different values of the cosmological matter density ($\Omega_m$) choices for $\Lambda$CDM cosmologies. Shown by the various lines are the ratios of the difference of each cosmology to those with a standard $\Omega_m=0.3$. Values of $\Omega_m$ are taken from Table~\ref{tab: omegas}}.
    \label{fig:enter-label}
\end{figure}

In Figure~\ref{fig:enter-label} we show how five different matter densities (motivated by their best fit values to the DESI, CMB and SN1a data - given in Table~\ref{tab: omegas}) affect the fractional number counts. We see that they have an interesting redshift dependence and can induce a sizeable difference (of order 15\%) compared to standard $\Omega_m = 0.3\, \Lambda$CDM. Of course, the values we have chosen correspond to the best fit $\Lambda$CDM values from the various experiments, and it is their difference from Planck that provides the driving force in the case of DESI for the need of dynamical dark energy to reconcile their results with those of Planck. The point however is that the number counts themselves provide a powerful discriminator for $\Omega_m$.

Ultimately we could show that many cosmological parameters would have an effect on the cosmological volume for high redshift galaxies. These effects all produce a change in the volume, and thus observed volume densities of galaxies, implying that a suite of possible cosmological parameters would need to be simulated into volumes using this framework and then to fit this volume evolution with the observational data to use this method as a way to measure evolving dark energy and the density parameter.

\begin{table}
\centering
\boxed{
\begin{tabular}{c|c}
\textbf{Model} & \textbf{$\Omega_m$}\\
\hline
DESI DR2 BAO     & 0.2975   \\
\hline
CMB              & 0.3169   \\
\hline
Pantheon+        & 0.331    \\
\hline
Union3           & 0.359    \\
\hline
DESY5            & 0.353   \\
\end{tabular}}
\caption{Values of $\Omega_m$ for different datasets - motivated by the best fit values obtained from the latest DESI results (see text).}
\label{tab: omegas}
\end{table}

\subsection{Future Surveys}

The above calculations have shown that in principle we will be able to reach a level of sensitivity sufficient to differentiate between several current models of evolving dark energy. 
The question then becomes: how much further must we improve our observations in order to reduce the statistical uncertainties to a level that meaningfully strengthens these constraints? We approach this by examining the Poisson uncertainties in galaxy number counts and the resulting uncertainties in the fitted cosmological parameters.

Because the Poisson uncertainty in the number counts propagates directly into uncertainties in parameters related to merger rates and, to a lesser extent, the star-formation rate, this provides a natural metric for estimating the required area of future surveys.

In a source-limited regime where galaxy counts follow Poisson statistics, the fractional uncertainty ($\sigma_N$) on the number of detected galaxies is, for a large measured number N,
\begin{equation}
    \left(\frac{\sigma_N}{N}\right) = \left(\frac{1}{\sqrt{N}}\right),
\end{equation}
with $N$ the number of galaxies in the survey area.  
If the surface density of galaxies is approximately uniform, then $N \propto A$, where $A$ is the angular area observed.  
The fractional Poisson uncertainty therefore scales as
\begin{equation}
    \left(\frac{\sigma_N}{N}\right) \propto \left(\frac{1}{\sqrt{A}}\right).
\end{equation}

\noindent To reduce the fractional uncertainty by a factor $f^{-1}\sim \epsilon$, the required survey area is
\begin{equation}
    A_{\mathrm{new}} = f^{2} A_{0},
\end{equation}
where $A_{0}$ is the initial survey area, which for our merger measurements is a combination of all the JWST deep fields for a total current area of $\sim 200$ arcmin$^{2}$ \citep[][]{Duan2024}. For our reference field of
\begin{equation}
    A_{0} = 189~\mathrm{arcmin}^{2} = 0.0525~\mathrm{deg}^{2},
\end{equation}
the areas required to reduce the Poisson fractional uncertainty by factors of $f = 100$ and $f = 1000$ are thus,
\begin{align}
    A_{100}  &= 100^{2} \times A_{0} = 1.89\times 10^{6}~\mathrm{arcmin}^{2}
               = 525~\mathrm{deg}^{2}, \\
    A_{1000} &= 1000^{2} \times A_{0} = 1.89\times 10^{8}~\mathrm{arcmin}^{2}
               = 5.25\times 10^{4}~\mathrm{deg}^{2}.
\end{align}

\noindent The requirement for a 1000-fold reduction exceeds the area of the entire sky
($A_{\rm sky} \simeq 41{,}253~\mathrm{deg}^{2}$), demonstrating that such a reduction cannot be achieved through increased survey area alone.  
By contrast, a 100-fold reduction (requiring $525~\mathrm{deg}^{2}$) is well within the scope of current and planned wide-area surveys. The \emph{Euclid} Wide Survey is expected to cover approximately $A_{\mathrm{Euclid}} \simeq 14{,}000\text{--}15{,}000~\mathrm{deg}^{2}$, while the Rubin Observatory LSST Wide--Fast--Deep (WFD) survey will image approximately $A_{\mathrm{LSST}} \simeq 18{,}000~\mathrm{deg}^{2}$ \citep[e.g.,][]{Mellier2024}.

The corresponding reductions in Poisson fractional uncertainty relative to our fiducial area are:
\begin{align}
    f_{\mathrm{Euclid}} &=
    \sqrt{\frac{A_{\mathrm{Euclid}}}{A_0}}
    \approx 516\text{--}535, \\
    f_{\mathrm{LSST}} &=
    \sqrt{\frac{A_{\mathrm{LSST}}}{A_0}}
    \approx 585.
\end{align}

\noindent With these large survey areas, both Euclid and LSST will dramatically reduce Poisson uncertainties compared to our initial 200~arcmin$^{2}$ area.  
The 100-fold reduction is easily surpassed, showing that this method can in principle work when measuring merger rates, and star-formation histories.  
While a full 1000-fold reduction would require slightly more than the full sky, the number counts can easily be increased by looking at fainter galaxies and reducing the limits of our integration.  This in some cases will require deeper depths than the next generation of surveys, but a careful consideration of depth and area would make it achievable in the near future.

However, it is clear that this method cannot be easily used today with the uncertainties on galaxy formation histories as measured through their star formation and merger rate.  The measurements of these quantities will improve dramatically with better modelling and measurements over the coming years. There are furthermore other approaches similar to this paper's which can be used to address this problem with different mass and redshift limits as well as fitting the various forms of evolution.  The coming years should see the advent of the data and models improved at a rate where this method will be as competitive as the leading cosmological methods today.

\section{Discussion and Summary} \label{sec:disc}

There exist a number of tried and tested observations used to constrain cosmological parameters. They include determining the best fitting model to the observed anisotropies in the cosmic microwave background radiation, the clustering of galaxies and associated weak lensing of light by the dark matter contained within the clusters, the observed rotation curves of individual galaxies, the large scale distribution of structure in the Universe and closer to home, the luminosity distance measurements to Cepheid variables and Type 1a Supernova in distant galaxies which have provided information on the rate of expansion of the universe, and most recently the direct detection of gravitational waves from the mergers of black hole binaries.  However, other ways to determine cosmological features and parameters are needed given the inconsistencies and tensions in various methods. 

The main goal of our paper is to describe a new method for determining the cosmological properties of the universe by using galaxy counts, coupled with the astrophysics of galaxy evolution.  It is motivated by a volume test, which was introduced in an early form by \cite{Tolman:1934za} in terms of galaxy counts,  but has never seriously been used because of the inherent difficulty associated with determining the key observables required to understand the evolution of  galaxies.  

The basic idea is that 
different cosmological models based on general relativity give us different volumes and thus, different number densities of galaxies which vary with cosmological properties. By measuring how galaxy counts change in volume bins over time we can in principle use this as a cosmological tool.    In an alternative form this method was used to show that the steady-state model of cosmology was inconsistent with the number counts of radio galaxies in the 1960s.  It was originally designed to determine the curvature of the universe, but can be applied to more general cosmological models. 

This idea has largely been abandoned for decades as it has been made clear since the work of \citep[e.g.,][]{Tinsley1968, Tinsley1980} that galaxies are dynamic -- they evolve in brightness, and their star formation/mergerrates and stellar masses change depending on redshift  \citep[e.g.,][]{Madau1998, Conselice2003,Harvey2024}.  However, it is clear that such is the pace of advance in galaxy observations, in particular with the JWST now probing out to large redshifts, and the impact future and ongoing surveys such as Euclid, LSST and Roman will have, the time is right to revisit this original idea. 

In this paper we have developed a new method to overcome these problems by accounting for the effects of galaxy evolution, in particular the increase in galaxy stellar mass over time due to star formation and galaxy mergers. We have shown a proof of principle calculation emphasizing that as the errors associated with these astrophysical quantities continue to decrease, galaxy number counts will become a competitive area for constraining the cosmological evolution of our universe.  

In particular in this paper we have investigated whether this volume test can be used to differentiate between a $\Lambda$CDM cosmology and a cosmology in which the universe is driven by dynamical dark energy.
By considering a particular paramerterisation of the dark energy equation of state, given in Eq.~\eqref{eq:eos-gen} with values for the parameters taken from Table~\ref{tab:gwmodels}, we have seen in Figure~\ref{fig: N merger comparison} that as the errors associated with the number counts decrease, we will be in a position where we can differentiate between a $\Lambda$CDM cosmology and a class of dynamical dark energy models.

These are at the limits of our knowledge of how to trace the evolution of dark energy and in the future we will want to differentiate between them.  
Our hope is that as
this method is further developed and applied in the future it will become as competitive as any other for measuring cosmological parameters and features.       


\section*{Acknowledgments}

CJC acknowledges support from the ERC Advanced Investigator Grant EPOCHS (788113). EJC acknowledges support from the STFC Consolidated Grant [ST/X000672/1] and also from a Leverhulme Research Fellowship [RF-2021 312]. SSM was supported by an STFC studentship [Grant No. ST/V506928/1], STFC Consolidated Grant No. ST/T001011/1 and by funds provided by the Center for Particle Cosmology at the University of Pennsylvania.

\section*{Data Availability}

There is no formal data used in this paper as it is based on analytical calculations and codes.   These codes and results in an electronic format will be shared on reasonable request to the authors.



\bibliographystyle{mnras}
\bibliography{main} 

@ARTICLE{Sandage1988,
       author = {{Sandage}, Allan},
        title = "{A Case for H 0 = 42 and Omega 0 = 1 Using Luminous Spiral Galaxies and the Cosmological Time Scale Test}",
      journal = {\apj},
         year = 1988,
        month = aug,
       volume = {331},
        pages = {583},
          doi = {10.1086/166584},
       adsurl = {https://ui.adsabs.harvard.edu/abs/1988ApJ...331..583S}
}

@ARTICLE{Casteels2014,
       author = {{Casteels}, Kevin R.~V. and {Conselice}, Christopher J. and {Bamford}, Steven P. and {Salvador-Sol{\'e}}, Eduard and {Norberg}, Peder R. and {Agius}, Nicola K. and {Baldry}, Ivan and {Brough}, Sarah and {Brown}, Michael J.~I. and {Drinkwater}, Michael J. and {Driver}, Simon P. and {Graham}, Alister W. and {Bland-Hawthorn}, Joss and {Hopkins}, Andrew M. and {Kelvin}, Lee S. and {L{\'o}pez-S{\'a}nchez}, Angel R. and {Loveday}, Jon and {Robotham}, Aaron S.~G. and {V{\'a}zquez-Mata}, Jos{\'e} A.},
        title = "{Galaxy And Mass Assembly (GAMA): refining the local galaxy merger rate using morphological information}",
      journal = {\mnras},
         year = 2014,
        month = dec,
       volume = {445},
       number = {2},
        pages = {1157-1169},
          doi = {10.1093/mnras/stu1799},
archivePrefix = {arXiv},
       eprint = {1409.1358},
 primaryClass = {astro-ph.GA},
       adsurl = {https://ui.adsabs.harvard.edu/abs/2014MNRAS.445.1157C}
}

@ARTICLE{Kaviraj2015,
       author = {{Kaviraj}, S. and {Devriendt}, J. and {Dubois}, Y. and {Slyz}, A. and {Welker}, C. and {Pichon}, C. and {Peirani}, S. and {Le Borgne}, D.},
        title = "{Galaxy merger histories and the role of merging in driving star formation at z > 1}",
      journal = {\mnras},
         year = 2015,
        month = sep,
       volume = {452},
       number = {3},
        pages = {2845-2850},
          doi = {10.1093/mnras/stv1500},
archivePrefix = {arXiv},
       eprint = {1411.2595},
 primaryClass = {astro-ph.GA},
       adsurl = {https://ui.adsabs.harvard.edu/abs/2015MNRAS.452.2845K}
}

@ARTICLE{Conselice2018,
       author = {{Conselice}, Christopher J. and {Twite}, Jonathan W. and {Palamara}, David P. and {Hartley}, William},
        title = "{The Halo Masses of Galaxies to z {\ensuremath{\sim}} 3: A Hybrid Observational and Theoretical Approach}",
      journal = {\apj},
         year = 2018,
        month = aug,
       volume = {863},
       number = {1},
          eid = {42},
        pages = {42},
          doi = {10.3847/1538-4357/aacda8},
archivePrefix = {arXiv},
       eprint = {1806.07752},
 primaryClass = {astro-ph.GA},
       adsurl = {https://ui.adsabs.harvard.edu/abs/2018ApJ...863...42C}
}

@ARTICLE{Mostar2013,
       author = {{Moster}, Benjamin P. and {Naab}, Thorsten and {White}, Simon D.~M.},
        title = "{Galactic star formation and accretion histories from matching galaxies to dark matter haloes}",
      journal = {\mnras},
         year = 2013,
        month = feb,
       volume = {428},
       number = {4},
        pages = {3121-3138},
          doi = {10.1093/mnras/sts261},
archivePrefix = {arXiv},
       eprint = {1205.5807},
 primaryClass = {astro-ph.CO},
       adsurl = {https://ui.adsabs.harvard.edu/abs/2013MNRAS.428.3121M}
}

@ARTICLE{Whitney2020,
       author = {{Whitney}, A. and {Conselice}, C.~J. and {Duncan}, K. and {Spitler}, L.~R.},
        title = "{Surface Brightness Evolution of Galaxies in the CANDELS GOODS Fields up to z {\ensuremath{\sim}} 6: High-z Galaxies Are Unique or Remain Undetected}",
      journal = {\apj},
         year = 2020,
        month = nov,
       volume = {903},
       number = {1},
          eid = {14},
        pages = {14},
          doi = {10.3847/1538-4357/abb824},
archivePrefix = {arXiv},
       eprint = {2009.07295},
 primaryClass = {astro-ph.GA},
       adsurl = {https://ui.adsabs.harvard.edu/abs/2020ApJ...903...14W}
}

@ARTICLE{Lofthouse2017,
       author = {{Lofthouse}, E.~K. and {Kaviraj}, S. and {Conselice}, C.~J. and {Mortlock}, A. and {Hartley}, W.},
        title = "{Major mergers are not significant drivers of star formation or morphological transformation around the epoch of peak cosmic star formation}",
      journal = {\mnras},
         year = 2017,
        month = mar,
       volume = {465},
       number = {3},
        pages = {2895-2900},
          doi = {10.1093/mnras/stw2895},
archivePrefix = {arXiv},
       eprint = {1608.03892},
 primaryClass = {astro-ph.GA},
       adsurl = {https://ui.adsabs.harvard.edu/abs/2017MNRAS.465.2895L}
}

@ARTICLE{DES2018,
       author = {{Abbott}, T.~M.~C. and {Abdalla}, F.~B. and {Alarcon}, A. and {Aleksi{\'c}}, J. and {Allam}, S. and {Allen}, S. and {Amara}, A. and {Annis}, J. and {Asorey}, J. and {Avila}, S. and {Bacon}, D. and {Balbinot}, E. and {Banerji}, M. and {Banik}, N. and {Barkhouse}, W. and {Baumer}, M. and {Baxter}, E. and {Bechtol}, K. and {Becker}, M.~R. and {Benoit-L{\'e}vy}, A. and {Benson}, B.~A. and {Bernstein}, G.~M. and {Bertin}, E. and {Blazek}, J. and {Bridle}, S.~L. and {Brooks}, D. and {Brout}, D. and {Buckley-Geer}, E. and {Burke}, D.~L. and {Busha}, M.~T. and {Campos}, A. and {Capozzi}, D. and {Carnero Rosell}, A. and {Carrasco Kind}, M. and {Carretero}, J. and {Castander}, F.~J. and {Cawthon}, R. and {Chang}, C. and {Chen}, N. and {Childress}, M. and {Choi}, A. and {Conselice}, C. and {Crittenden}, R. and {Crocce}, M. and {Cunha}, C.~E. and {D'Andrea}, C.~B. and {da Costa}, L.~N. and {Das}, R. and {Davis}, T.~M. and {Davis}, C. and {De Vicente}, J. and {DePoy}, D.~L. and {DeRose}, J. and {Desai}, S. and {Diehl}, H.~T. and {Dietrich}, J.~P. and {Dodelson}, S. and {Doel}, P. and {Drlica-Wagner}, A. and {Eifler}, T.~F. and {Elliott}, A.~E. and {Elsner}, F. and {Elvin-Poole}, J. and {Estrada}, J. and {Evrard}, A.~E. and {Fang}, Y. and {Fernandez}, E. and {Fert{\'e}}, A. and {Finley}, D.~A. and {Flaugher}, B. and {Fosalba}, P. and {Friedrich}, O. and {Frieman}, J. and {Garc{\'\i}a-Bellido}, J. and {Garcia-Fernandez}, M. and {Gatti}, M. and {Gaztanaga}, E. and {Gerdes}, D.~W. and {Giannantonio}, T. and {Gill}, M.~S.~S. and {Glazebrook}, K. and {Goldstein}, D.~A. and {Gruen}, D. and {Gruendl}, R.~A. and {Gschwend}, J. and {Gutierrez}, G. and {Hamilton}, S. and {Hartley}, W.~G. and {Hinton}, S.~R. and {Honscheid}, K. and {Hoyle}, B. and {Huterer}, D. and {Jain}, B. and {James}, D.~J. and {Jarvis}, M. and {Jeltema}, T. and {Johnson}, M.~D. and {Johnson}, M.~W.~G. and {Kacprzak}, T. and {Kent}, S. and {Kim}, A.~G. and {King}, A. and {Kirk}, D. and {Kokron}, N. and {Kovacs}, A. and {Krause}, E. and {Krawiec}, C. and {Kremin}, A. and {Kuehn}, K. and {Kuhlmann}, S. and {Kuropatkin}, N. and {Lacasa}, F. and {Lahav}, O. and {Li}, T.~S. and {Liddle}, A.~R. and {Lidman}, C. and {Lima}, M. and {Lin}, H. and {MacCrann}, N. and {Maia}, M.~A.~G. and {Makler}, M. and {Manera}, M. and {March}, M. and {Marshall}, J.~L. and {Martini}, P. and {McMahon}, R.~G. and {Melchior}, P. and {Menanteau}, F. and {Miquel}, R. and {Miranda}, V. and {Mudd}, D. and {Muir}, J. and {M{\"o}ller}, A. and {Neilsen}, E. and {Nichol}, R.~C. and {Nord}, B. and {Nugent}, P. and {Ogando}, R.~L.~C. and {Palmese}, A. and {Peacock}, J. and {Peiris}, H.~V. and {Peoples}, J. and {Percival}, W.~J. and {Petravick}, D. and {Plazas}, A.~A. and {Porredon}, A. and {Prat}, J. and {Pujol}, A. and {Rau}, M.~M. and {Refregier}, A. and {Ricker}, P.~M. and {Roe}, N. and {Rollins}, R.~P. and {Romer}, A.~K. and {Roodman}, A. and {Rosenfeld}, R. and {Ross}, A.~J. and {Rozo}, E. and {Rykoff}, E.~S. and {Sako}, M. and {Salvador}, A.~I. and {Samuroff}, S. and {S{\'a}nchez}, C. and {Sanchez}, E. and {Santiago}, B. and {Scarpine}, V. and {Schindler}, R. and {Scolnic}, D. and {Secco}, L.~F. and {Serrano}, S. and {Sevilla-Noarbe}, I. and {Sheldon}, E. and {Smith}, R.~C. and {Smith}, M. and {Smith}, J. and {Soares-Santos}, M. and {Sobreira}, F. and {Suchyta}, E. and {Tarle}, G. and {Thomas}, D. and {Troxel}, M.~A. and {Tucker}, D.~L. and {Tucker}, B.~E. and {Uddin}, S.~A. and {Varga}, T.~N. and {Vielzeuf}, P. and {Vikram}, V. and {Vivas}, A.~K. and {Walker}, A.~R. and {Wang}, M. and {Wechsler}, R.~H. and {Weller}, J. and {Wester}, W. and {Wolf}, R.~C. and {Yanny}, B. and {Yuan}, F. and {Zenteno}, A. and {Zhang}, B. and {Zhang}, Y. and {Zuntz}, J.},
        title = "{Dark Energy Survey year 1 results: Cosmological constraints from galaxy clustering and weak lensing}",
      journal = {\prd},
         year = 2018,
        month = aug,
       volume = {98},
       number = {4},
          eid = {043526},
        pages = {043526},
          doi = {10.1103/PhysRevD.98.043526},
archivePrefix = {arXiv},
       eprint = {1708.01530},
 primaryClass = {astro-ph.CO},
       adsurl = {https://ui.adsabs.harvard.edu/abs/2018PhRvD..98d3526A}
}

@article{Mundy2015,
    title = {{Tracing galaxy populations through cosmic time: a critical test of methods for connecting the same galaxies between different redshifts at z < 3}},
    year = {2015},
    journal = {Monthly Notices of the Royal Astronomical Society},
    author = {Mundy, C. J. and Conselice, C. J. and Ownsworth, J. R.},
    number = {4},
    month = {5},
    pages = {3696--3707},
    volume = {450},
    url = {http://adsabs.harvard.edu/abs/2015MNRAS.450.3696M http://mnras.oxfordjournals.org/cgi/doi/10.1093/mnras/stv860},
    doi = {10.1093/mnras/stv860},
    issn = {0035-8711}
}

@ARTICLE{Mellier2024,
       author = {{Euclid Collaboration} and {Mellier}, Y. and {Abdurro'uf} and {Acevedo Barroso}, J.~A. and {Ach{\'u}carro}, A. and {Adamek}, J. and {Adam}, R. and {Addison}, G.~E. and {Aghanim}, N. and {Aguena}, M. and {Ajani}, V. and {Akrami}, Y. and {Al-Bahlawan}, A. and {Alavi}, A. and {Albuquerque}, I.~S. and {Alestas}, G. and {Alguero}, G. and {Allaoui}, A. and {Allen}, S.~W. and {Allevato}, V. and {Alonso-Tetilla}, A.~V. and {Altieri}, B. and {Alvarez-Candal}, A. and {Alvi}, S. and {Amara}, A. and {Amendola}, L. and {Amiaux}, J. and {Andika}, I.~T. and {Andreon}, S. and {Andrews}, A. and {Angora}, G. and {Angulo}, R.~E. and {Annibali}, F. and {Anselmi}, A. and {Anselmi}, S. and {Arcari}, S. and {Archidiacono}, M. and {Aric{\`o}}, G. and {Arnaud}, M. and {Arnouts}, S. and {Asgari}, M. and {Asorey}, J. and {Atayde}, L. and {Atek}, H. and {Atrio-Barandela}, F. and {Aubert}, M. and {Aubourg}, E. and {Auphan}, T. and {Auricchio}, N. and {Aussel}, B. and {Aussel}, H. and {Avelino}, P.~P. and {Avgoustidis}, A. and {Avila}, S. and {Awan}, S. and {Azzollini}, R. and {Baccigalupi}, C. and {Bachelet}, E. and {Bacon}, D. and {Baes}, M. and {Bagley}, M.~B. and {Bahr-Kalus}, B. and {Balaguera-Antolinez}, A. and {Balbinot}, E. and {Balcells}, M. and {Baldi}, M. and {Baldry}, I. and {Balestra}, A. and {Ballardini}, M. and {Ballester}, O. and {Balogh}, M. and {Ba{\~n}ados}, E. and {Barbier}, R. and {Bardelli}, S. and {Baron}, M. and {Barreiro}, T. and {Barrena}, R. and {Barriere}, J. -C. and {Barros}, B.~J. and {Barthelemy}, A. and {Bartolo}, N. and {Basset}, A. and {Battaglia}, P. and {Battisti}, A.~J. and {Baugh}, C.~M. and {Baumont}, L. and {Bazzanini}, L. and {Beaulieu}, J. -P. and {Beckmann}, V. and {Belikov}, A.~N. and {Bel}, J. and {Bellagamba}, F. and {Bella}, M. and {Bellini}, E. and {Benabed}, K. and {Bender}, R. and {Benevento}, G. and {Bennett}, C.~L. and {Benson}, K. and {Bergamini}, P. and {Bermejo-Climent}, J.~R. and {Bernardeau}, F. and {Bertacca}, D. and {Berthe}, M. and {Berthier}, J. and {Bethermin}, M. and {Beutler}, F. and {Bevillon}, C. and {Bhargava}, S. and {Bhatawdekar}, R. and {Bianchi}, D. and {Bisigello}, L. and {Biviano}, A. and {Blake}, R.~P. and {Blanchard}, A. and {Blazek}, J. and {Blot}, L. and {Bosco}, A. and {Bodendorf}, C. and {Boenke}, T. and {B{\"o}hringer}, H. and {Boldrini}, P. and {Bolzonella}, M. and {Bonchi}, A. and {Bonici}, M. and {Bonino}, D. and {Bonino}, L. and {Bonvin}, C. and {Bon}, W. and {Booth}, J.~T. and {Borgani}, S. and {Borlaff}, A.~S. and {Borsato}, E. and {Bosco}, A. and {Bose}, B. and {Botticella}, M.~T. and {Boucaud}, A. and {Bouche}, F. and {Boucher}, J.~S. and {Boutigny}, D. and {Bouvard}, T. and {Bouwens}, R. and {Bouy}, H. and {Bowler}, R.~A.~A. and {Bozza}, V. and {Bozzo}, E. and {Branchini}, E. and {Brando}, G. and {Brau-Nogue}, S. and {Brekke}, P. and {Bremer}, M.~N. and {Brescia}, M. and {Breton}, M. -A. and {Brinchmann}, J. and {Brinckmann}, T. and {Brockley-Blatt}, C. and {Brodwin}, M. and {Brouard}, L. and {Brown}, M.~L. and {Bruton}, S. and {Bucko}, J. and {Buddelmeijer}, H. and {Buenadicha}, G. and {Buitrago}, F. and {Burger}, P. and {Burigana}, C. and {Busillo}, V. and {Busonero}, D. and {Cabanac}, R. and {Cabayol-Garcia}, L. and {Cagliari}, M.~S. and {Caillat}, A. and {Caillat}, L. and {Calabrese}, M. and {Calabro}, A. and {Calderone}, G. and {Calura}, F. and {Camacho Quevedo}, B. and {Camera}, S. and {Campos}, L. and {Canas-Herrera}, G. and {Candini}, G.~P. and {Cantiello}, M. and {Capobianco}, V. and {Cappellaro}, E. and {Cappelluti}, N. and {Cappi}, A. and {Caputi}, K.~I. and {Cara}, C. and {Carbone}, C. and {Cardone}, V.~F. and {Carella}, E. and {Carlberg}, R.~G. and {Carle}, M. and {Carminati}, L. and {Caro}, F. and {Carrasco}, J.~M. and {Carretero}, J. and {Carrilho}, P. and {Carron Duque}, J.},
        title = "{Euclid. I. Overview of the Euclid mission}",
      journal = {arXiv e-prints},
         year = 2024,
        month = may,
          eid = {arXiv:2405.13491},
        pages = {arXiv:2405.13491},
          doi = {10.48550/arXiv.2405.13491},
archivePrefix = {arXiv},
       eprint = {2405.13491},
 primaryClass = {astro-ph.CO},
       adsurl = {https://ui.adsabs.harvard.edu/abs/2024arXiv240513491E}
}

@ARTICLE{duncan2019,
       author = {{Duncan}, Kenneth and {Conselice}, Christopher J. and {Mundy}, Carl and {Bell}, Eric and {Donley}, Jennifer and {Galametz}, Audrey and {Guo}, Yicheng and {Grogin}, Norman A. and {Hathi}, Nimish and {Kartaltepe}, Jeyhan and {Kocevski}, Dale and {Koekemoer}, Anton M. and {P{\'e}rez-Gonz{\'a}lez}, Pablo G. and {Mantha}, Kameswara B. and {Snyder}, Gregory F. and {Stefanon}, Mauro},
        title = "{Observational Constraints on the Merger History of Galaxies since z {\ensuremath{\approx}} 6: Probabilistic Galaxy Pair Counts in the CANDELS Fields}",
      journal = {\apj},
         year = 2019,
        month = may,
       volume = {876},
       number = {2},
          eid = {110},
        pages = {110},
          doi = {10.3847/1538-4357/ab148a},
archivePrefix = {arXiv},
       eprint = {1903.12188},
 primaryClass = {astro-ph.GA},
       adsurl = {https://ui.adsabs.harvard.edu/abs/2019ApJ...876..110D}
}

@ARTICLE{Ferreira2020,
       author = {{Ferreira}, Leonardo and {Conselice}, Christopher J. and {Duncan}, Kenneth and {Cheng}, Ting-Yun and {Griffiths}, Alex and {Whitney}, Amy},
        title = "{Galaxy Merger Rates up to z {\ensuremath{\sim}} 3 Using a Bayesian Deep Learning Model: A Major-merger Classifier Using IllustrisTNG Simulation Data}",
      journal = {\apj},
         year = 2020,
        month = jun,
       volume = {895},
       number = {2},
          eid = {115},
        pages = {115},
          doi = {10.3847/1538-4357/ab8f9b},
archivePrefix = {arXiv},
       eprint = {2005.00476},
 primaryClass = {astro-ph.GA},
       adsurl = {https://ui.adsabs.harvard.edu/abs/2020ApJ...895..115F}
}

@ARTICLE{Duan2024,
       author = {{Duan}, Qiao and {Conselice}, Christopher J. and {Li}, Qiong and {Austin}, Duncan and {Harvey}, Thomas and {Adams}, Nathan J. and {Duncan}, Kenneth J. and {Trussler}, James and {Ferreira}, Leonardo and {Westcott}, Lewi and {Harris}, Honor and {Windhorst}, Rogier A. and {Holwerda}, Benne W. and {Broadhurst}, Thomas J. and {Coe}, Dan and {Cohen}, Seth H. and {Driver}, Simon P. and {Frye}, Brenda and {Grogin}, Norman A. and {Hathi}, Nimish P. and {Jansen}, Rolf A. and {Koekemoer}, Anton M. and {Marshall}, Madeline A. and {Nonino}, Mario and {Ortiz}, Rafael, III and {Pirzkal}, Nor and {Robotham}, Aaron and {Ryan}, Russell E., Jr and {Summers}, Jake and {D'Silva}, Jordan C.~J. and {Willmer}, Christopher N.~A. and {Yan}, Haojing},
        title = "{Galaxy Mergers in the Epoch of Reionization I: A JWST Study of Pair Fractions, Merger Rates, and Stellar Mass Accretion Rates at $z = 4.5-11.5$}",
      journal = {arXiv e-prints},
         year = 2024,
        month = jul,
          eid = {arXiv:2407.09472},
        pages = {arXiv:2407.09472},
          doi = {10.48550/arXiv.2407.09472},
archivePrefix = {arXiv},
       eprint = {2407.09472},
 primaryClass = {astro-ph.GA},
       adsurl = {https://ui.adsabs.harvard.edu/abs/2024arXiv240709472D}
}

@ARTICLE{shibuya2022,
       author = {{Shibuya}, Takatoshi and {Miura}, Noriaki and {Iwadate}, Kenji and {Fujimoto}, Seiji and {Harikane}, Yuichi and {Toba}, Yoshiki and {Umayahara}, Takuya and {Ito}, Yohito},
        title = "{Galaxy morphologies revealed with Subaru HSC and super-resolution techniques. I. Major merger fractions of L$_{UV}$   3-15 L*$_{UV}$ dropout galaxies at z   4-7}",
      journal = {\pasj},
         year = 2022,
        month = feb,
       volume = {74},
       number = {1},
        pages = {73-91},
          doi = {10.1093/pasj/psab111},
archivePrefix = {arXiv},
       eprint = {2106.03728},
 primaryClass = {astro-ph.GA},
       adsurl = {https://ui.adsabs.harvard.edu/abs/2022PASJ...74...73S}
}

@ARTICLE{husko2022,
       author = {{Hu{\v{s}}ko}, Filip and {Lacey}, Cedric G. and {Baugh}, Carlton M.},
        title = "{Statistics of galaxy mergers: bridging the gap between theory and observation}",
      journal = {\mnras},
         year = 2022,
        month = feb,
       volume = {509},
       number = {4},
        pages = {5918-5937},
          doi = {10.1093/mnras/stab3324},
archivePrefix = {arXiv},
       eprint = {2107.05601},
 primaryClass = {astro-ph.GA},
       adsurl = {https://ui.adsabs.harvard.edu/abs/2022MNRAS.509.5918H}
}

@ARTICLE{Mantha2018,
       author = {{Mantha}, Kameswara Bharadwaj and {McIntosh}, Daniel H. and {Brennan}, Ryan and {Ferguson}, Henry C. and {Kodra}, Dritan and {Newman}, Jeffrey A. and {Rafelski}, Marc and {Somerville}, Rachel S. and {Conselice}, Christopher J. and {Cook}, Joshua S. and {Hathi}, Nimish P. and {Koo}, David C. and {Lotz}, Jennifer M. and {Simmons}, Brooke D. and {Straughn}, Amber N. and {Snyder}, Gregory F. and {Wuyts}, Stijn and {Bell}, Eric F. and {Dekel}, Avishai and {Kartaltepe}, Jeyhan and {Kocevski}, Dale D. and {Koekemoer}, Anton M. and {Lee}, Seong-Kook and {Lucas}, Ray A. and {Pacifici}, Camilla and {Peth}, Michael A. and {Barro}, Guillermo and {Dahlen}, Tomas and {Finkelstein}, Steven L. and {Fontana}, Adriano and {Galametz}, Audrey and {Grogin}, Norman A. and {Guo}, Yicheng and {Mobasher}, Bahram and {Nayyeri}, Hooshang and {P{\'e}rez-Gonz{\'a}lez}, Pablo G. and {Pforr}, Janine and {Santini}, Paola and {Stefanon}, Mauro and {Wiklind}, Tommy},
        title = "{Major merging history in CANDELS. I. Evolution of the incidence of massive galaxy-galaxy pairs from z = 3 to z {\ensuremath{\sim}} 0}",
      journal = {\mnras},
         year = 2018,
        month = apr,
       volume = {475},
       number = {2},
        pages = {1549-1573},
          doi = {10.1093/mnras/stx3260},
archivePrefix = {arXiv},
       eprint = {1712.06611},
 primaryClass = {astro-ph.GA},
       adsurl = {https://ui.adsabs.harvard.edu/abs/2018MNRAS.475.1549M}
}

@article{Bluck2012,
    title = {{THE STRUCTURES AND TOTAL (MINOR + MAJOR) MERGER HISTORIES OF MASSIVE GALAXIES UP TO z ∼ 3 IN THE HST GOODS NICMOS SURVEY: A POSSIBLE SOLUTION TO THE SIZE EVOLUTION PROBLEM}},
    year = {2012},
    journal = {The Astrophysical Journal},
    author = {Bluck, Asa F. L. and Conselice, Christopher J. and Buitrago, Fernando and Gr{\"{u}}tzbauch, Ruth and Hoyos, Carlos and Mortlock, Alice and Bauer, Amanda E.},
    number = {1},
    month = {3},
    pages = {34},
    volume = {747},
    url = {http://adsabs.harvard.edu/abs/2012ApJ...747...34B http://stacks.iop.org/0004-637X/747/i=1/a=34?key=crossref.f1f2fc21fe0959a2e22c4b6ff43b6e4b},
    doi = {10.1088/0004-637X/747/1/34},
    issn = {0004-637X}
}

@article{Bluck2009,
    title = {{A surprisingly high pair fraction for extremely massive galaxies at z ∼ 3 in the GOODS NICMOS survey}},
    year = {2009},
    journal = {Monthly Notices of the Royal Astronomical Society: Letters},
    author = {Bluck, Asa F. L. and Conselice, Christopher J. and Bouwens, Rychard J. and Daddi, Emanuele and Dickinson, Mark and Papovich, Casey and Yan, Haojing},
    number = {1},
    month = {3},
    pages = {L51-L55},
    volume = {394},
    url = {http://adsabs.harvard.edu/abs/2009MNRAS.394L..51B http://mnrasl.oxfordjournals.org/cgi/doi/10.1111/j.1745-3933.2008.00608.x},
    doi = {10.1111/j.1745-3933.2008.00608.x},
    issn = {17453925}
}

@article{Conselice2003,
    title = {{A Direct Measurement of Major Galaxy Mergers at z 3}},
    year = {2003},
    journal = {The Astronomical Journal},
    author = {Conselice, Christopher J. and Bershady, Matthew A. and Dickinson, Mark and Papovich, Casey},
    number = {3},
    month = {9},
    pages = {1183--1207},
    volume = {126},
    url = {http://adsabs.harvard.edu/abs/2003AJ....126.1183C http://arxiv.org/abs/astro-ph/0306106 http://stacks.iop.org/1538-3881/126/i=3/a=1183},
    isbn = {0004-6256},
    doi = {10.1086/377318},
    issn = {0004-6256},
    arxivId = {astro-ph/0306106},
    language = {en}
}

@ARTICLE{Adams2023,
       author = {{Adams}, Nathan J. and {Conselice}, Christopher J. and {Austin}, Duncan and {Harvey}, Thomas and {Ferreira}, Leonardo and {Trussler}, James and {Juodzbalis}, Ignas and {Li}, Qiong and {Windhorst}, Rogier and {Cohen}, Seth H. and {Jansen}, Rolf and {Summers}, Jake and {Tompkins}, Scott and {Driver}, Simon P. and {Robotham}, Aaron and {D'Silva}, Jordan C.~J. and {Yan}, Haojing and {Coe}, Dan and {Frye}, Brenda and {Grogin}, Norman A. and {Koekemoer}, Anton M. and {Marshall}, Madeline A. and {Pirzkal}, Nor and {Ryan}, Russell E., Jr. and {Maksym}, W. Peter and {Rutkowski}, Michael J. and {Willmer}, Christopher N.~A. and {Hammel}, Heidi B. and {Nonino}, Mario and {Bhatawdekar}, Rachana and {Wilkins}, Stephen M. and {Willner}, Steven P. and {Bradley}, Larry D. and {Broadhurst}, Tom and {Cheng}, Cheng and {Dole}, Herve and {Hathi}, Nimish P. and {Zitrin}, Adi},
        title = "{EPOCHS Paper II: The Ultraviolet Luminosity Function from $7.5<z<13.5$ using 110 square arcminutes of deep, blank-field data from the PEARLS Survey and Public Science Programmes}",
      journal = {arXiv e-prints},
         year = 2023,
        month = apr,
          eid = {arXiv:2304.13721},
        pages = {arXiv:2304.13721},
          doi = {10.48550/arXiv.2304.13721},
archivePrefix = {arXiv},
       eprint = {2304.13721},
 primaryClass = {astro-ph.GA},
       adsurl = {https://ui.adsabs.harvard.edu/abs/2023arXiv230413721A}
}

@article{Frieman:2008sn,
    author = "Frieman, Joshua and Turner, Michael and Huterer, Dragan",
    title = "{Dark Energy and the Accelerating Universe}",
    eprint = "0803.0982",
    archivePrefix = "arXiv",
    primaryClass = "astro-ph",
    reportNumber = "FERMILAB-PUB-08-613-A",
    doi = "10.1146/annurev.astro.46.060407.145243",
    journal = "Ann. Rev. Astron. Astrophys.",
    volume = "46",
    pages = "385--432",
    year = "2008"
}

@article{Amendola:2016saw,
    author = "Amendola, Luca and others",
    title = "{Cosmology and fundamental physics with the Euclid satellite}",
    eprint = "1606.00180",
    archivePrefix = "arXiv",
    primaryClass = "astro-ph.CO",
    doi = "10.1007/s41114-017-0010-3",
    journal = "Living Rev. Rel.",
    volume = "21",
    number = "1",
    pages = "2",
    year = "2018"
}

@article{Bundy:2005kd,
    author = "Bundy, Kevin and Ellis, Richard S. and Conselice, Christopher J. and Taylor, James E. and Cooper, Michael C. and Willmer, Christopher N. A. and Weiner, Benjamin J. and Noeske, Kai G. and Eisendardt, Peter R. M.",
    title = "{The mass assembly history of field galaxies: detection of an evolving mass limit for star forming galaxies}",
    eprint = "astro-ph/0512465",
    archivePrefix = "arXiv",
    doi = "10.1086/507456",
    journal = "Astrophys. J.",
    volume = "651",
    pages = "120--141",
    year = "2006"
}

@ARTICLE{ryle1961,
       author = {{Ryle}, M. and {Clarke}, R.~W.},
        title = "{An examination of the steady-state model in the light of some recent observations of radio sources}",
      journal = {\mnras},
         year = 1961,
        month = jan,
       volume = {122},
        pages = {349},
          doi = {10.1093/mnras/122.4.349},
       adsurl = {https://ui.adsabs.harvard.edu/abs/1961MNRAS.122..349R}
}

@ARTICLE{Mortlock2015,
       author = {{Mortlock}, Alice and {Conselice}, Christopher. J. and {Hartley}, William G. and {Duncan}, Ken and {Lani}, Caterina and {Ownsworth}, Jamie R. and {Almaini}, Omar and {Wel}, Arjen van der and {Huang}, Kuang-Han and {Ashby}, Matthew L.~N. and {Willner}, S.~P. and {Fontana}, Adriano and {Dekel}, Avishai and {Koekemoer}, Anton M. and {Ferguson}, Harry C. and {Faber}, Sandra M. and {Grogin}, Norman A. and {Kocevski}, Dale D.},
        title = "{Deconstructing the galaxy stellar mass function with UKIDSS and CANDELS: the impact of colour, structure and environment}",
      journal = {\mnras},
         year = 2015,
        month = feb,
       volume = {447},
       number = {1},
        pages = {2-24},
          doi = {10.1093/mnras/stu2403},
archivePrefix = {arXiv},
       eprint = {1411.3339},
 primaryClass = {astro-ph.GA},
       adsurl = {https://ui.adsabs.harvard.edu/abs/2015MNRAS.447....2M}
}

@ARTICLE{Yoshii1988A,
       author = {{Yoshii}, Yuzuru and {Takahara}, Fumio},
        title = "{Galactic Evolution and Cosmology: Probing the Cosmological Deceleration Parameter}",
      journal = {\apj},
         year = 1988,
        month = mar,
       volume = {326},
        pages = {1},
          doi = {10.1086/166065},
       adsurl = {https://ui.adsabs.harvard.edu/abs/1988ApJ...326....1Y}
}

@ARTICLE{HubbleTolman1935,
       author = {{Hubble}, Edwin and {Tolman}, Richard C.},
        title = "{Two Methods of Investigating the Nature of the Nebular Redshift}",
      journal = {\apj},
         year = 1935,
        month = nov,
       volume = {82},
        pages = {302},
          doi = {10.1086/143682},
       adsurl = {https://ui.adsabs.harvard.edu/abs/1935ApJ....82..302H}
}

@ARTICLE{Conselice2014morph,
       author = {{Conselice}, Christopher J.},
        title = "{The Evolution of Galaxy Structure Over Cosmic Time}",
      journal = {\araa},
         year = 2014,
        month = aug,
       volume = {52},
        pages = {291-337},
          doi = {10.1146/annurev-astro-081913-040037},
archivePrefix = {arXiv},
       eprint = {1403.2783},
 primaryClass = {astro-ph.GA},
       adsurl = {https://ui.adsabs.harvard.edu/abs/2014ARA&A..52..291C}
}

@article{Conselice:2016zid,
    author = "Conselice, Christopher J. and Wilkinson, Aaron and Duncan, Kenneth and Mortlock, Alice",
    title = "{THE EVOLUTION OF GALAXY NUMBER DENSITY ATz\ensuremath{<} 8 AND ITS IMPLICATIONS}",
    doi = "10.3847/0004-637x/830/2/83",
    journal = "Astrophys. J.",
    volume = "830",
    number = "2",
    pages = "83",
    year = "2016"
}

@article{Copeland:2006wr,
    author = "Copeland, Edmund J. and Sami, M. and Tsujikawa, Shinji",
    title = "{Dynamics of dark energy}",
    eprint = "hep-th/0603057",
    archivePrefix = "arXiv",
    doi = "10.1142/S021827180600942X",
    journal = "Int. J. Mod. Phys. D",
    volume = "15",
    pages = "1753--1936",
    year = "2006"
}

@article{Duncan2014,
  title={The mass evolution of the first galaxies: stellar mass functions and star formation rates at 4< z< 7 in the CANDELS GOODS-South field},
  author={Duncan, Kenneth and Conselice, Christopher J and Mortlock, Alice and Hartley, William G and Guo, Yicheng and Ferguson, Henry C and Dav{\'e}, Romeel and Lu, Yu and Ownsworth, Jamie and Ashby, Matthew LN and others},
  journal={Monthly Notices of the Royal Astronomical Society},
  volume={444},
  number={3},
  pages={2960--2984},
  year={2014},
  publisher={The Royal Astronomical Society}
}

@article{Freedman:2023jcz,
    author = "Freedman, Wendy L. and Madore, Barry F.",
    title = "{Progress in direct measurements of the Hubble constant}",
    eprint = "2309.05618",
    archivePrefix = "arXiv",
    primaryClass = "astro-ph.CO",
    doi = "10.1088/1475-7516/2023/11/050",
    journal = "JCAP",
    volume = "11",
    pages = "050",
    year = "2023"
}

@article{Hounsell:2017ejq,
    author = "Hounsell, R. and others",
    title = "{Simulations of the WFIRST Supernova Survey and Forecasts of Cosmological Constraints}",
    eprint = "1702.01747",
    archivePrefix = "arXiv",
    primaryClass = "astro-ph.IM",
    reportNumber = "FERMILAB-PUB-17-075-PPD",
    doi = "10.3847/1538-4357/aac08b",
    journal = "Astrophys. J.",
    volume = "867",
    number = "1",
    pages = "23",
    year = "2018"
}

@article{Kamionkowski:2022pkx,
    author = "Kamionkowski, Marc and Riess, Adam G.",
    title = "{The Hubble Tension and Early Dark Energy}",
    eprint = "2211.04492",
    archivePrefix = "arXiv",
    primaryClass = "astro-ph.CO",
    journal = "Ann. Rev. Nucl. Part. Sci.",
    volume = "73",
    pages = "153--180",
    year = "2023"
}

@article{Knox:2019rjx,
    author = "Knox, Lloyd and Millea, Marius",
    title = "{Hubble constant hunter\textquoteright{}s guide}",
    eprint = "1908.03663",
    archivePrefix = "arXiv",
    primaryClass = "astro-ph.CO",
    doi = "10.1103/PhysRevD.101.043533",
    journal = "Phys. Rev. D",
    volume = "101",
    number = "4",
    pages = "043533",
    year = "2020"
}

@article{Larson:2010gs,
    author = "Larson, D. and others",
    title = "{Seven-Year Wilkinson Microwave Anisotropy Probe (WMAP) Observations: Power Spectra and WMAP-Derived Parameters}",
    eprint = "1001.4635",
    archivePrefix = "arXiv",
    primaryClass = "astro-ph.CO",
    doi = "10.1088/0067-0049/192/2/16",
    journal = "Astrophys. J. Suppl.",
    volume = "192",
    pages = "16",
    year = "2011"
}

@article{Loh:1986wg,
    author = "Loh, E. D. and Spillar, E. J.",
    title = "{A measurement of the mass density of the universe}",
    doi = "10.1086/184717",
    journal = "Astrophys. J. Lett.",
    volume = "307",
    pages = "L1",
    year = "1986"
}

@article{LSST:2008ijt,
    author = "Ivezi\'c, \v{Z}eljko and others",
    collaboration = "LSST",
    title = "{LSST: from Science Drivers to Reference Design and Anticipated Data Products}",
    eprint = "0805.2366",
    archivePrefix = "arXiv",
    primaryClass = "astro-ph",
    reportNumber = "SLAC-PUB-16076",
    doi = "10.3847/1538-4357/ab042c",
    journal = "Astrophys. J.",
    volume = "873",
    number = "2",
    pages = "111",
    year = "2019"
}

@article{Madau2014,
   author = {Piero Madau and Mark Dickinson},
   doi = {10.1146/annurev-astro-081811-125615},
   issn = {0066-4146, 1545-4282},
   issue = {1},
   journal = {Annual Review of Astronomy and Astrophysics},
   month = {8},
   pages = {415-486},
   title = {Cosmic Star-Formation History},
   volume = {52},
   url = {https://www.annualreviews.org/doi/10.1146/annurev-astro-081811-125615},
   year = {2014},
}

@article{Madau1998,
   author = {Piero Madau and Lucia Pozzetti and Mark Dickinson},
   doi = {10.1086/305523},
   issn = {0004-637X},
   issue = {1},
   journal = {The Astrophysical Journal},
   month = {5},
   pages = {106-116},
   title = {The Star Formation History of Field Galaxies},
   volume = {498},
   year = {1998},
}

@article{Madore:1991yf,
    author = "Madore, Barry F. and Freedman, Wendy L.",
    title = "{The Cepheid distance scale}",
    doi = "10.1086/132911",
    journal = "Publ. Astron. Soc. Pac.",
    volume = "103",
    pages = "933--957",
    year = "1991"
}

@article{Planck:2018vyg,
    author = "Aghanim, N. and others",
    collaboration = "Planck",
    title = "{Planck 2018 results. VI. Cosmological parameters}",
    eprint = "1807.06209",
    archivePrefix = "arXiv",
    primaryClass = "astro-ph.CO",
    doi = "10.1051/0004-6361/201833910",
    journal = "Astron. Astrophys.",
    volume = "641",
    pages = "A6",
    year = "2020",
    note = "[Erratum: Astron.Astrophys. 652, C4 (2021)]"
}

@article{SupernovaSearchTeam:1998fmf,
    author = "Riess, Adam G. and others",
    collaboration = "Supernova Search Team",
    title = "{Observational evidence from supernovae for an accelerating universe and a cosmological constant}",
    eprint = "astro-ph/9805201",
    archivePrefix = "arXiv",
    doi = "10.1086/300499",
    journal = "Astron. J.",
    volume = "116",
    pages = "1009--1038",
    year = "1998"
}

@article{SupernovaCosmologyProject:1998vns,
    author = "Perlmutter, S. and others",
    collaboration = "Supernova Cosmology Project",
    title = "{Measurements of $\Omega$ and $\Lambda$ from 42 High Redshift Supernovae}",
    eprint = "astro-ph/9812133",
    archivePrefix = "arXiv",
    reportNumber = "LBNL-41801, LBL-41801",
    doi = "10.1086/307221",
    journal = "Astrophys. J.",
    volume = "517",
    pages = "565--586",
    year = "1999"
}

@article{Tolman:1934za,
    author = "Tolman, Richard C.",
    title = "{Effect of imhomogeneity on cosmological models}",
    doi = "10.1073/pnas.20.3.169",
    journal = "Proc. Nat. Acad. Sci.",
    volume = "20",
    pages = "169--176",
    year = "1934"
}

@ARTICLE{Tinsley1968,
       author = {{Tinsley}, Beatrice M.},
        title = "{Evolution of the Stars and Gas in Galaxies}",
      journal = {\apj},
         year = 1968,
        month = feb,
       volume = {151},
        pages = {547},
          doi = {10.1086/149455},
       adsurl = {https://ui.adsabs.harvard.edu/abs/1968ApJ...151..547T}
}

@ARTICLE{Baldy2012,
       author = {{Baldry}, I.~K. and {Driver}, S.~P. and {Loveday}, J. and {Taylor}, E.~N. and {Kelvin}, L.~S. and {Liske}, J. and {Norberg}, P. and {Robotham}, A.~S.~G. and {Brough}, S. and {Hopkins}, A.~M. and {Bamford}, S.~P. and {Peacock}, J.~A. and {Bland-Hawthorn}, J. and {Conselice}, C.~J. and {Croom}, S.~M. and {Jones}, D.~H. and {Parkinson}, H.~R. and {Popescu}, C.~C. and {Prescott}, M. and {Sharp}, R.~G. and {Tuffs}, R.~J.},
        title = "{Galaxy And Mass Assembly (GAMA): the galaxy stellar mass function at z < 0.06}",
      journal = {\mnras},
         year = 2012,
        month = mar,
       volume = {421},
       number = {1},
        pages = {621-634},
          doi = {10.1111/j.1365-2966.2012.20340.x},
archivePrefix = {arXiv},
       eprint = {1111.5707},
 primaryClass = {astro-ph.CO},
       adsurl = {https://ui.adsabs.harvard.edu/abs/2012MNRAS.421..621B}
}

@ARTICLE{Ownsworth2016,
       author = {{Ownsworth}, Jamie R. and {Conselice}, Christopher J. and {Mundy}, Carl J. and {Mortlock}, Alice and {Hartley}, William G. and {Duncan}, Kenneth and {Almaini}, Omar},
        title = "{The evolution of galaxies at constant number density: a less biased view of star formation, quenching, and structural formation}",
      journal = {\mnras},
         year = 2016,
        month = sep,
       volume = {461},
       number = {1},
        pages = {1112-1129},
          doi = {10.1093/mnras/stw1207},
archivePrefix = {arXiv},
       eprint = {1605.05741},
 primaryClass = {astro-ph.GA},
       adsurl = {https://ui.adsabs.harvard.edu/abs/2016MNRAS.461.1112O}
}

@ARTICLE{Whitney2021,
       author = {{Whitney}, A. and {Ferreira}, L. and {Conselice}, C.~J. and {Duncan}, K.},
        title = "{Galaxy Evolution in All Five CANDELS Fields and IllustrisTNG: Morphological, Structural, and the Major Merger Evolution to z   3}",
      journal = {\apj},
         year = 2021,
        month = oct,
       volume = {919},
       number = {2},
          eid = {139},
        pages = {139},
          doi = {10.3847/1538-4357/ac1422},
archivePrefix = {arXiv},
       eprint = {2105.01675},
 primaryClass = {astro-ph.GA},
       adsurl = {https://ui.adsabs.harvard.edu/abs/2021ApJ...919..139W}
}

@ARTICLE{Conselice2013,
       author = {{Conselice}, Christopher J. and {Mortlock}, Alice and {Bluck}, Asa F.~L. and {Gr{\"u}tzbauch}, Ruth and {Duncan}, Kenneth},
        title = "{Gas accretion as a dominant formation mode in massive galaxies from the GOODS NICMOS Survey}",
      journal = {\mnras},
         year = 2013,
        month = apr,
       volume = {430},
       number = {2},
        pages = {1051-1060},
          doi = {10.1093/mnras/sts682},
archivePrefix = {arXiv},
       eprint = {1206.6995},
 primaryClass = {astro-ph.CO},
       adsurl = {https://ui.adsabs.harvard.edu/abs/2013MNRAS.430.1051C}
}

@article{ACT:2020gnv,
    author = "Aiola, Simone and others",
    collaboration = "ACT",
    title = "{The Atacama Cosmology Telescope: DR4 Maps and Cosmological Parameters}",
    eprint = "2007.07288",
    archivePrefix = "arXiv",
    primaryClass = "astro-ph.CO",
    doi = "10.1088/1475-7516/2020/12/047",
    journal = "JCAP",
    volume = "12",
    pages = "047",
    year = "2020"
}

@ARTICLE{Conselice2008,
       author = {{Conselice}, Christopher J. and {Rajgor}, Sheena and {Myers}, Robert},
        title = "{The structures of distant galaxies - I. Galaxy structures and the merger rate to z \raisebox{-0.5ex}\textasciitilde 3 in the Hubble Ultra-Deep Field}",
      journal = {\mnras},
         year = 2008,
        month = may,
       volume = {386},
       number = {2},
        pages = {909-927},
          doi = {10.1111/j.1365-2966.2008.13069.x},
archivePrefix = {arXiv},
       eprint = {0711.2333},
 primaryClass = {astro-ph},
       adsurl = {https://ui.adsabs.harvard.edu/abs/2008MNRAS.386..909C}
}

@ARTICLE{Tinsley1980,
       author = {{Tinsley}, B.~M.},
        title = "{Evolution of the Stars and Gas in Galaxies}",
      journal = {\fcp},
         year = 1980,
        month = jan,
       volume = {5},
        pages = {287-388},
          doi = {10.48550/arXiv.2203.02041},
archivePrefix = {arXiv},
       eprint = {2203.02041},
 primaryClass = {astro-ph.GA},
       adsurl = {https://ui.adsabs.harvard.edu/abs/1980FCPh....5..287T}
}

@ARTICLE{Austin2026,
       author = {{Austin}, Duncan and {Harvey}, Thomas and {Conselice}, Christopher J. and {Adams}, Nathan J. and {Rusakov}, Vadim and {Li}, Qiong and {Westcott}, Lewi and {Goolsby}, Caio and {Madgwick}, Kai and {Arcidiacono}, James and {Ricotti}, Massimo and {Newman}, Sophie L. and {Seeyave}, Louise T.~C. and {Trussler}, James and {Frye}, Brenda and {Grogin}, Norman A. and {Jansen}, Rolf A. and {Koekemoer}, Anton M. and {Pirzkal}, Nor and {Rutkowski}, Michael and {Windhorst}, Rogier A.},
        title = "{Resolving the ionizing photon budget crisis with JWST/NIRCam HII clumping constraints at z=6}",
      journal = {arXiv e-prints},
         year = 2025,
        month = dec,
          eid = {arXiv:2512.10839},
        pages = {arXiv:2512.10839},
          doi = {10.48550/arXiv.2512.10839},
archivePrefix = {arXiv},
       eprint = {2512.10839},
 primaryClass = {astro-ph.GA},
       adsurl = {https://ui.adsabs.harvard.edu/abs/2025arXiv251210839A}
}

@ARTICLE{Ellis1997,
       author = {{Ellis}, Richard S.},
        title = "{Faint Blue Galaxies}",
      journal = {\araa},
         year = 1997,
        month = jan,
       volume = {35},
        pages = {389-443},
          doi = {10.1146/annurev.astro.35.1.389},
archivePrefix = {arXiv},
       eprint = {astro-ph/9704019},
 primaryClass = {astro-ph},
       adsurl = {https://ui.adsabs.harvard.edu/abs/1997ARA&A..35..389E}
}

@ARTICLE{Broadhurst,
       author = {{Broadhurst}, T.~J. and {Ellis}, R.~S. and {Shanks}, T.},
        title = "{The Durham/Anglo-Australian telescope faint galaxy redshift survey.}",
      journal = {\mnras},
         year = 1988,
        month = dec,
       volume = {235},
        pages = {827-856},
          doi = {10.1093/mnras/235.3.827},
       adsurl = {https://ui.adsabs.harvard.edu/abs/1988MNRAS.235..827B}
}

@ARTICLE{Steidel1996,
       author = {{Steidel}, Charles C. and {Giavalisco}, Mauro and {Pettini}, Max and {Dickinson}, Mark and {Adelberger}, Kurt L.},
        title = "{Spectroscopic Confirmation of a Population of Normal Star-forming Galaxies at Redshifts Z > 3}",
      journal = {\apjl},
         year = 1996,
        month = may,
       volume = {462},
        pages = {L17},
          doi = {10.1086/310029},
archivePrefix = {arXiv},
       eprint = {astro-ph/9602024},
 primaryClass = {astro-ph},
       adsurl = {https://ui.adsabs.harvard.edu/abs/1996ApJ...462L..17S}
}

@ARTICLE{Williams1996,
       author = {{Williams}, Robert E. and {Blacker}, Brett and {Dickinson}, Mark and {Dixon}, W. Van Dyke and {Ferguson}, Henry C. and {Fruchter}, Andrew S. and {Giavalisco}, Mauro and {Gilliland}, Ronald L. and {Heyer}, Inge and {Katsanis}, Rocio and {Levay}, Zolt and {Lucas}, Ray A. and {McElroy}, Douglas B. and {Petro}, Larry and {Postman}, Marc and {Adorf}, Hans-Martin and {Hook}, Richard},
        title = "{The Hubble Deep Field: Observations, Data Reduction, and Galaxy Photometry}",
      journal = {\aj},
         year = 1996,
        month = oct,
       volume = {112},
        pages = {1335},
          doi = {10.1086/118105},
archivePrefix = {arXiv},
       eprint = {astro-ph/9607174},
 primaryClass = {astro-ph},
       adsurl = {https://ui.adsabs.harvard.edu/abs/1996AJ....112.1335W}
}

@ARTICLE{Shanks1984,
       author = {{Shanks}, T. and {Stevenson}, P.~R.~F. and {Fong}, R. and {MacGillivray}, H.~T.},
        title = "{Galaxy number counts and cosmology}",
      journal = {\mnras},
         year = 1984,
        month = feb,
       volume = {206},
        pages = {767-800},
          doi = {10.1093/mnras/206.4.767},
       adsurl = {https://ui.adsabs.harvard.edu/abs/1984MNRAS.206..767S}
}

@BOOK{Peebles1980,
       author = {{Peebles}, P.~J.~E.},
        title = "{The large-scale structure of the universe}",
         year = 1980,
       adsurl = {https://ui.adsabs.harvard.edu/abs/1980lssu.book.....P}
}

@INPROCEEDINGS{Cowie1991,
       author = {{Cowie}, L.~L.},
        title = "{K Band Galaxy Counts and the Cosmological Geometry}",
    booktitle = {Observational Tests of Cosmological Inflation},
         year = 1991,
       editor = {{Shanks}, T. and {Banday}, A.~J. and {Ellis}, R.~S.},
       series = {NATO Advanced Study Institute (ASI) Series C},
       volume = {348},
        month = jan,
        pages = {257},
       adsurl = {https://ui.adsabs.harvard.edu/abs/1991ASIC..348..257C}
}

@ARTICLE{Harvey2024,
       author = {{Harvey}, Thomas and {Conselice}, Christopher J. and {Adams}, Nathan J. and {Austin}, Duncan and {Juod{\v{z}}balis}, Ignas and {Trussler}, James and {Li}, Qiong and {Ormerod}, Katherine and {Ferreira}, Leonardo and {Lovell}, Christopher C. and {Duan}, Qiao and {Westcott}, Lewi and {Harris}, Honor and {Bhatawdekar}, Rachana and {Coe}, Dan and {Cohen}, Seth H. and {Caruana}, Joseph and {Cheng}, Cheng and {Driver}, Simon P. and {Frye}, Brenda and {Furtak}, Lukas J. and {Grogin}, Norman A. and {Hathi}, Nimish P. and {Holwerda}, Benne W. and {Jansen}, Rolf A. and {Koekemoer}, Anton M. and {Marshall}, Madeline A. and {Nonino}, Mario and {Vijayan}, Aswin P. and {Wilkins}, Stephen M. and {Windhorst}, Rogier and {Willmer}, Christopher N.~A. and {Yan}, Haojing and {Zitrin}, Adi},
        title = "{EPOCHS. IV. SED Modeling Assumptions and Their Impact on the Stellar Mass Function at 6.5 {\ensuremath{\leq}} z {\ensuremath{\leq}} 13.5 Using PEARLS and Public JWST Observations}",
      journal = {\apj},
         year = 2025,
        month = jan,
       volume = {978},
       number = {1},
          eid = {89},
        pages = {89},
          doi = {10.3847/1538-4357/ad8c29},
archivePrefix = {arXiv},
       eprint = {2403.03908},
 primaryClass = {astro-ph.GA},
       adsurl = {https://ui.adsabs.harvard.edu/abs/2025ApJ...978...89H}
}

@article{DESI:2024mwx,
    author = "Adame, A. G. and others",
    collaboration = "DESI",
    title = "{DESI 2024 VI: cosmological constraints from the measurements of baryon acoustic oscillations}",
    eprint = "2404.03002",
    archivePrefix = "arXiv",
    primaryClass = "astro-ph.CO",
    reportNumber = "FERMILAB-PUB-24-0154-PPD",
    doi = "10.1088/1475-7516/2025/02/021",
    journal = "JCAP",
    volume = "02",
    pages = "021",
    year = "2025"
}

@article{Cortes:2024lgw,
    author = "Cort\^es, Marina and Liddle, Andrew R.",
    title = "{Interpreting DESI's evidence for evolving dark energy}",
    eprint = "2404.08056",
    archivePrefix = "arXiv",
    primaryClass = "astro-ph.CO",
    doi = "10.1088/1475-7516/2024/12/007",
    journal = "JCAP",
    volume = "12",
    pages = "007",
    year = "2024",

}

@article{Cortes:2025joz,
    author = "Cort\^es, Marina and Liddle, Andrew R.",
    title = "{On DESI's DR2 exclusion of $\Lambda$CDM}",
    eprint = "2504.15336",
    archivePrefix = "arXiv",
    primaryClass = "astro-ph.CO",
    month = "4",
    year = "2025",
    journal = {arXiv e-prints},
    eprint = "2503.14738",
    adsurl = {https://ui.adsabs.harvard.edu/abs/2025arXiv250415336C/abstract},
    archivePrefix = "arXiv",
       eid = {2504.15336},
        pages = {2504.15336},
          doi = {10.48550/arXiv.2504.15336}
}

@ARTICLE{1988ApJ...326L...1B,
       author = {{Bahcall}, Safi R. and {Tremaine}, Scott},
        title = "{Evolutionary Corrections to the Redshift-Volume Measurement of the Density Parameter}",
      journal = {\apjl},
         year = 1988,
        month = mar,
       volume = {326},
        pages = {L1},
          doi = {10.1086/185110},
       adsurl = {https://ui.adsabs.harvard.edu/abs/1988ApJ...326L...1B}
}

@article{pacifici2016timing,
  title={Timing the Evolution of Quiescent and Star-forming Local Galaxies},
  author={Pacifici, Camilla and Oh, Sree and Oh, Kyuseok and Lee, Jaehyun and Sukyoung, K Yi},
  journal={The Astrophysical Journal},
  volume={824},
  number={1},
  pages={45},
  year={2016},
  publisher={IOP Publishing}
}

@ARTICLE{2000ApJ...534L..11N,
       author = {{Newman}, Jeffrey A. and {Davis}, Marc},
        title = "{Measuring the Cosmic Equation of State with Counts of Galaxies}",
      journal = {\apjl},
         year = 2000,
        month = may,
       volume = {534},
       number = {1},
        pages = {L11-L14},
          doi = {10.1086/312657},
archivePrefix = {arXiv},
       eprint = {astro-ph/9912366},
 primaryClass = {astro-ph},
       adsurl = {https://ui.adsabs.harvard.edu/abs/2000ApJ...534L..11N}
}

@article{schechter1976analytic,
  title={An analytic expression for the luminosity function for galaxies.},
  author={Schechter, Paul},
  journal={Astrophysical Journal, Vol. 203, p. 297-306},
  volume={203},
  pages={297--306},
  year={1976}
}

@article{Linder_2003,
   title={Exploring the Expansion History of the Universe},
   volume={90},
   ISSN={1079-7114},
   url={http://dx.doi.org/10.1103/PhysRevLett.90.091301},
   DOI={10.1103/physrevlett.90.091301},
   number={9},
   journal={Physical Review Letters},
   publisher={American Physical Society (APS)},
   author={Linder, Eric V.},
   year={2003},
   month=mar }

@article{Efstathiou:2024xcq,
    author = "Efstathiou, George",
    title = "{Evolving dark energy or supernovae systematics?}",
    eprint = "2408.07175",
    archivePrefix = "arXiv",
    primaryClass = "astro-ph.CO",
    doi = "10.1093/mnras/staf301",
    journal = "Mon. Not. Roy. Astron. Soc.",
    volume = "538",
    number = "2",
    pages = "875--882",
    year = "2025"
}

@article{Wolf:2025jed,
    author = "Wolf, William J. and Garc\'\i{}a-Garc\'\i{}a, Carlos and Anton, Theodore and Ferreira, Pedro G.",
    title = "{Assessing cosmological evidence for non-minimal coupling}",
    eprint = "2504.07679",
    archivePrefix = "arXiv",
    primaryClass = "astro-ph.CO",
    month = "4",
    year = "2025",
    journal = {arXiv e-prints},
    eprint = "2504.07679",
    adsurl = {https://ui.adsabs.harvard.edu/abs/2025arXiv250407679W/abstract},
    archivePrefix = "arXiv",
       eid = {2504.07679},
        pages = {2504.07679},
          doi = {10.48550/arXiv.2504.07679}
}

@article{Shah:2020jmc,
    author = "Shah, Ekta A. and others",
    title = "{Investigating the Effect of Galaxy Interactions on the Enhancement of Active Galactic Nuclei at 0.5 \ensuremath{<} $z$ \ensuremath{<} 3.0}",
    eprint = "2010.02710",
    archivePrefix = "arXiv",
    primaryClass = "astro-ph.GA",
    doi = "10.3847/1538-4357/abbf59",
    journal = "Astrophys. J.",
    volume = "904",
    number = "2",
    pages = "107",
    year = "2020"
}

@article{Man:2014jsa,
    author = "Man, Allison W. S. and Zirm, Andrew W. and Toft, Sune",
    title = "{Resolving the Discrepancy of Galaxy Merger Fraction Measurements at z \textasciitilde{} 0 - 3}",
    eprint = "1410.3479",
    archivePrefix = "arXiv",
    primaryClass = "astro-ph.GA",
    doi = "10.3847/0004-637X/830/2/89",
    journal = "Astrophys. J.",
    volume = "830",
    number = "2",
    pages = "89",
    year = "2016"
}

@article{Mundy:2015qgi,
    author = {Mundy, Carl J. and Conselice, Christopher J. and Duncan, Kenneth J. and Almaini, Omar and H\"au\ss{}ler, Boris and Hartley, William G.},
    title = "{A consistent measure of the merger histories of massive galaxies using close-pair statistics \textendash{} I. Major mergers at z \ensuremath{<} 3.5}",
    eprint = "1705.07986",
    archivePrefix = "arXiv",
    primaryClass = "astro-ph.GA",
    doi = "10.1093/mnras/stx1238",
    journal = "Mon. Not. Roy. Astron. Soc.",
    volume = "450",
    pages = "3696--3707",
    year = "2015"
}

@article{DESI:2025zgx,
    author = "Abdul Karim, M. and others",
    collaboration = "DESI",
    title = "{DESI DR2 Results II: Measurements of Baryon Acoustic Oscillations and Cosmological Constraints}",
    journal = {arXiv e-prints},
    eprint = "2503.14738",
    adsurl = {https://ui.adsabs.harvard.edu/abs/2025arXiv250314738D/abstract},
    archivePrefix = "arXiv",
       eid = {arXiv:2503.14738},
        pages = {arXiv:2503.14738},
          doi = {10.48550/arXiv.2503.14738},
    primaryClass = "astro-ph.CO",
    month = "3",
    year = "2025"
}

@ARTICLE{2025arXiv251023823H,
       author = {{H0DN Collaboration} and {Casertano}, Stefano and {Anand}, Gagandeep and {Anderson}, Richard I. and {Beaton}, Rachael and {Bhardwaj}, Anupam and {Blakeslee}, John P. and {Boubel}, Paula and {Breuval}, Louise and {Brout}, Dillon and {Cantiello}, Michele and {Cruz Reyes}, Mauricio and {Cs{\"o}rnyei}, Geza and {de Jaeger}, Thomas and {Dhawan}, Suhail and {Di Valentino}, Eleonora and {Galbany}, Llu{\'\i}s and {Gil-Mar{\'\i}n}, H{\'e}ctor and {Graczyk}, Dariusz and {Huang}, Caroline and {Jensen}, Joseph B. and {Kervella}, Pierre and {Leibundgut}, Bruno and {Lengen}, Bastian and {Li}, Siyang and {Macri}, Lucas and {{\"O}z{\"u}lker}, Emre and {Pesce}, Dominic W. and {Riess}, Adam and {Romaniello}, Martino and {Said}, Khaled and {Sch{\"o}neberg}, Nils and {Scolnic}, Dan and {Sicignano}, Teresa and {Skowron}, Dorota M. and {Uddin}, Syed A. and {Verde}, Licia and {Nota}, Antonella},
        title = "{The Local Distance Network: a community consensus report on the measurement of the Hubble constant at 1\% precision}",
      journal = {arXiv e-prints},
         year = 2025,
        month = oct,
          eid = {arXiv:2510.23823},
        pages = {arXiv:2510.23823},
          doi = {10.48550/arXiv.2510.23823},
archivePrefix = {arXiv},
       eprint = {2510.23823},
 primaryClass = {astro-ph.CO},
       adsurl = {https://ui.adsabs.harvard.edu/abs/2025arXiv251023823H}
}






    \label{fig: Gridplot1}



\bsp	
\label{lastpage}
\end{document}